\documentclass[final,5p,times,twocolumn,authoryear]{elsarticle}
\usepackage{amssymb}
\usepackage{amsmath}
\usepackage{lmodern}
\usepackage{subcaption}
\usepackage{lineno}
\usepackage{caption}
\usepackage{float}
\usepackage[most]{tcolorbox}
\tcbuselibrary{listings, breakable}
\usepackage{cuted}
\tcbset{
	algobox/.style={
		colback=gray!5,
		colframe=black!50,
		fonttitle=\bfseries,
		title=Algorithm: PEFRF (Permutation Entropy-Based Features + Random Forest),
		boxrule=0.4pt,
		arc=1mm,
		top=1mm, bottom=1mm,
		left=1mm, right=1mm,
		enhanced,
		breakable
	}
}
\usepackage{tabularx}
\newcolumntype{C}[1]{>{\centering\arraybackslash}p{#1}}
\usepackage{setspace}
\usepackage{array}
\usepackage{multirow}
\usepackage{booktabs}
\usepackage{colortbl}
\usepackage{xcolor}
\usepackage{caption}
\usepackage{graphicx}
\usepackage{adjustbox}

\usepackage{siunitx}
\usepackage{geometry}
\makeatletter
\newcommand\sotiny{\@setfontsize\sotiny\@vpt\@vpt}
\newcommand\ntiny{\@setfontsize\notsotiny\@vpt\@viipt}
\makeatother
\definecolor{uclagold}{rgb}{1.0, 0.7, 0.0}
\definecolor{bubblegum}{rgb}{0.99, 0.76, 0.8}
\definecolor{cyan(process)}{rgb}{0.6, 0.9, 0.9}
\definecolor{lime(colorwheel)}{rgb}{0.75, 1.0, 0.0}
\definecolor{lavender(floral)}{rgb}{0.71, 0.49, 0.86}
\definecolor{unmyellow}{rgb}{0.93, 0.91, 0.67}
\definecolor{grey}{rgb}{0.7, 0.75, 0.71}

\newcommand\crule[3][black]{\textcolor{#1}{\rule{#2}{#3}}}

\journal{Computer Vision and Image Understanding}

\begin{document}

\begin{frontmatter}



\title{A Hybrid Approach for Unified Image Quality Assessment: Permutation Entropy-Based Features Fused with Random Forest for Natural-Scene and Screen-Content Images for Cross-Content Applications}
\author[label1]{Mohtashim Baqar} 
\author[label2]{Sian Lun Lau}  
\author[label3]{Mansoor Ebrahim} 
\affiliation[label1]{organization={School of IT and Engineering, Melbourne Institute of Technology},
            addressline={\\154-158 Sussex Street}, 
            city={Sydney},
            postcode={2000}, 
            state={New South Wales},
            country={Australia}}
\affiliation[label2]{organization={School of Computing and Artificial Intelligence, Faculty of Engineering and Technology, Sunway University},
			addressline={\\5, Jalan Universiti}, 
			city={Bandar Sunway},
			postcode={47500}, 
			state={Selangor},
			country={Malaysia}}
\affiliation[label3]{organization={Faculty of Engineering, Sciences, and Technology, Iqra University (Main Campus)},
			city={\\Karachi},
			postcode={75500}, 
			state={Sindh},
			country={Pakistan}}
\begin{abstract}
Image Quality Assessment (IQA) plays a vital role in applications such as image compression, restoration, and multimedia streaming. However, existing metrics often struggle to generalize across diverse image types—particularly between natural-scene images (NSIs) and screen-content images (SCIs)—due to their differing structural and perceptual characteristics. To address this limitation, we propose a novel full-reference IQA framework: Permutation Entropy-based Features Fused with Random Forest (PEFRF). PEFRF captures structural complexity by extracting permutation entropy from the gradient maps of reference, distorted, and fused images, forming a robust feature vector. These features are then input into a Random Forest regressor trained on subjective quality scores to predict final image quality. The framework is evaluated on 13 benchmark datasets comprising over 21,000 images and 40+ state-of-the-art IQA metrics. Experimental results demonstrate that PEFRF consistently outperforms existing methods across various distortion types and content domains, establishing its effectiveness as a unified and statistically significant solution for cross-content image quality assessment.
\end{abstract}

\end{frontmatter}
\section{Introduction}
Ensuring high-quality visual experiences is a critical goal of modern multimedia systems, especially as user expectations grow in domains such as image compression, video streaming, augmented reality (AR), and medical imaging. At the core of this objective lies Image Quality Assessment (IQA), which serves as a fundamental tool for evaluating and optimizing image fidelity.\par 
Traditional IQA metrics, such as MSE, PSNR, and SSIM, have been widely adopted in industry and research. However, their limitations are increasingly exposed in emerging applications involving dynamic, heterogeneous, or interactive content. For instance, in AR systems, where visual content continuously changes in response to user interaction, traditional metrics often fail to reflect perceptual degradation accurately—leading to reduced user immersion and usability, as noted by Duan \citep{duan2022confusing}.\par 
Compounding this issue is the exponential growth of multimedia content, which spans a wide range of image types. Among these, natural-scene images (NSIs) and screen-content images (SCIs) represent two distinct categories. While NSIs feature natural textures and lighting variations, SCIs—such as computer graphics, medical scans, and screen recordings—exhibit sharper edges, uniform backgrounds, and high-frequency structures. These differences severely challenge the generalizability of traditional IQA methods.\par 
To address these limitations, we propose a unified full-reference IQA framework called PEFRF (Permutation Entropy-based Features Fused with Random Forest). The method combines permutation entropy-based structural features with a machine learning-based regressor to enable content-agnostic, perceptually aligned image quality assessment. Unlike conventional approaches tailored to specific image types or distortion categories, PEFRF is designed to generalize across both natural-scene and screen-content images.
\subsection{Challenges in Image Quality Assessment}
Despite continuous progress in image quality assessment (IQA), several critical challenges hinder the development of truly robust and generalizable frameworks. These challenges span perceptual, computational, and application-specific dimensions \citep{wang2016objective}:
\begin{itemize}
	\item \textbf{Viewing Conditions and Display Settings:} Perceived image quality varies significantly based on factors such as viewing distance, screen resolution, brightness, and ambient lighting. Metrics that do not account for these variations may yield inconsistent results across devices and settings.
	\item \textbf{Context-Dependent Sensitivity:} Human perception is more sensitive to distortions in semantically important regions (e.g., faces, text) than in background textures. This creates difficulty in designing IQA models that align closely with subjective judgments.
	\item \textbf{Complex and Mixed Distortions:} Real-world images often suffer from overlapping degradation types (e.g., compression, blur, noise), which complicates the assessment process. Most traditional models are optimized for single distortions.
	\item \textbf{Mismatch Between Subjective and Objective Quality:} Human perception is inherently subjective, influenced by cognitive factors and prior experience. Bridging the gap between objective metric outputs and subjective ratings remains a persistent challenge.
	\item \textbf{Neglect of Temporal Artefacts in Videos:} Many IQA models are static in nature and fail to consider temporal factors such as flickering, motion blur, or frame drops, which critically affect quality in video sequences.
	\item \textbf{Limited Dataset Diversity:} Existing IQA benchmarks often focus on natural-scene images and overlook domains like satellite imagery, medical scans, or computer graphics. This restricts the generalization ability of IQA models.
	\item \textbf{Device-Specific Rendering Variations:} Perception of image quality can differ based on device characteristics (e.g., OLED vs. LCD, HDR vs. SDR, desktop vs. mobile), but many IQA models assume a uniform viewing environment.
	\item \textbf{Trade-off Between Accuracy and Efficiency:} While hand-crafted metrics such as SSIM are computationally light, they lack fine perceptual accuracy. Conversely, deep learning-based models offer higher accuracy but are resource-intensive.
	\item \textbf{Susceptibility to Adversarial Perturbations:} Deep neural network-based IQA models can be manipulated by imperceptible image modifications, raising concerns about their robustness in security-sensitive applications.
	\item \textbf{Limited Cross-Domain Generalization:} Models trained on natural image datasets often underperform when applied to domains like medical imaging, thermal vision, or synthetic imagery due to content-specific biases.
\end{itemize}
\section{Background and Related Work}
The field of Image Quality Assessment (IQA) has witnessed significant advancements, yet challenges persist in developing unified frameworks capable of accurately evaluating diverse image content types. Recent studies have attempted to bridge this gap, offering insights into the complexities of IQA methodologies. 
\subsection{Limitations --- Existing IQA Approaches}
A significant limitation of existing IQA methods lies in their dependency on specific content types and distortion characteristics. For instance, traditional FR approaches like Multi-Scale Structural Similarity (MS-SSIM) and Gradient Magnitude Similarity Deviation (GMSD) perform well on natural scene images but struggle with screen-content images, such as medical scans or computer-generated graphics. Similarly, NR methods trained on natural image datasets fail to generalize to screen-content images, where distortions have unique statistical properties \citep{chandler2014seven}.\par 
Moreover, many existing IQA frameworks inadequately model human perception, especially for screen-content images. While entropy-based metrics, such as those using permutation entropy, provide a promising approach to capture image complexity and information density, their stand-alone application has shown limited effectiveness in capturing fine-grained distortions or adapting to mixed-content datasets. Additionally, machine learning-based IQA methods, such as convolutional neural network-based approaches, while effective, often suffer from high computational complexity, requiring extensive training and large-scale annotated datasets. This limitation restricts their practical deployment in real-time or resource-constrained environments, such as video streaming or embedded systems.\par 
Further, existing attempts at unified IQA frameworks have their own limitations. For example, \citep{min2017unified} introduces a Unified Content-Type Adaptive (UCA) model for Blind Image Quality Assessment (BIQA) of Compressed Natural Scene Images (NSIs), Computer-Generated Images (CGIs), and Screen Content Images (SCIs). While the model effectively addresses challenges posed by varying content types and compression distortions using a novel cross-content-type database and adaptive multi-scale weighting, it is primarily designed for block-based compression methods (e.g., HEVC, HEVC-SCC). This focus limits its applicability to other common distortion types, such as noise and blur. Additionally, the UCA model's weighting mechanism applies a single adaptive strategy based on overall content likelihood without segmenting the image into distinct content-specific regions, resulting in suboptimal performance for highly heterogeneous images.
\subsection{Recent Advances in IQA}
Recent studies have explored the integration of machine learning techniques with perceptual features to enhance IQA performance. For example, the authors leverage Random Forest (RA) regression to build an IQA model that aligns closely with human perception. By using features derived from Difference of Gaussian (DoG) frequency bands, which mimic the Human Visual System (HVS), the Random Forest framework ensures robust and accurate predictions across diverse IQA databases \citep{pei2015image}. \par 
Furthermore, the exploration of content fidelity in NR-IQA has been addressed by Zhou et al. \citep{zhou2025blind}, who proposed a quality adversarial learning framework emphasizing both content fidelity and prediction accuracy. This approach dynamically adapts and refines the image quality assessment process based on quality optimization results, highlighting the importance of content fidelity in NR-IQA. 
\subsection{Identified Gaps in Literature}
From the review of existing methods, the following critical gaps are identified:
\begin{itemize}
	\item Generalization Across Content Types: Existing frameworks often focus on either NSIs or SCIs, failing to develop a unified approach capable of handling both content types effectively.
	\item Perceptual Feature Integration: While entropy-based features are promising, their potential remains under explored, particularly in combination with ensemble learning models like Random Forests.
	\item Scalability and Efficiency: The computational overhead of many IQA methods limits their practicality, particularly for applications requiring real-time or resource-efficient solutions. 
\end{itemize}
In summary, existing IQA models suffer from limited adaptability to diverse content types, especially when applied across both NSIs and SCIs. While some recent efforts have attempted to bridge this gap, they either compromise on perceptual accuracy or lack computational scalability. This gap highlights the need for a unified, content-agnostic IQA approach that balances generalization, efficiency, and interpretability. These insights form the basis for our proposed framework, described in the following section.
\subsection{Unified IQA Frameworks}
Yun and Lin \citep{yun2023uniqa} introduced the "You Only Train Once" (YOTO) framework, aiming to unify Full-Reference (FR) and No-Reference (NR) IQA tasks. By employing an encoder to extract multi-level features and integrating a Hierarchical Attention module alongside a Semantic Distortion Aware module, YOTO adapts to both FR and NR inputs. The framework demonstrated state-of-the-art performance when trained independently on NR or FR tasks and showed enhanced NR IQA performance when trained jointly on both tasks. However, YOTO primarily focuses on the methodological unification of FR and NR assessments rather than addressing the diverse nature of image content types. \par 
Zhou et al. \citep{zhou2024uniaa} proposed the Unified Multi-modal Image Aesthetic Assessment (UNIAA) framework, leveraging Multi-modal Large Language Models (MLLMs) to align image aesthetic assessment with human perception. Despite its innovative approach, UNIAA is primarily centred on aesthetic evaluation and may not fully capture the technical quality aspects pertinent to diverse image content.
\subsection{Theoretical Motivation and Proposed Framework}
This section outlines the theoretical basis and conceptual motivations behind the proposed PEFRF framework. We first identify the inherent challenges in achieving unified IQA across diverse content types, followed by an exploration of permutation entropy as a perceptually meaningful complexity measure. These insights culminate in the design rationale behind the PEFRF architecture.
\subsection{Challenges in Unified IQA}
\begin{itemize} 
	\item Algorithm Limitations: Current state-of-the-art (SOTA) methods in Image Quality Assessment (IQA) are often optimized for specific content types, such as natural scene images (NSIs). These methods exhibit a marked decline in performance when applied to screen-content images (SCIs), computer-generated graphics, or mixed-modality content, due to the distinctive statistical and perceptual characteristics inherent to SCIs \citep{8000398}. This discrepancy arises because traditional frameworks lack the ability to effectively capture and adapt to the content-specific features of SCIs, highlighting a significant gap in algorithmic adaptability.
	\item Generalization Gaps: Despite the increasing diversity of visual data in the real world, relatively few studies \citep{min2017unified} have focused on developing unified IQA frameworks capable of handling mixed content types seamlessly. Existing approaches typically rely on hand-crafted or learned features optimized for a single content domain, which limits their ability to generalize across heterogeneous datasets. This limitation is particularly problematic in practical scenarios, such as multimedia platforms or cross-domain applications, where images of varying origins and characteristics coexist. 
\end{itemize}
These challenges underscore a fundamental limitation in current IQA methodologies: the absence of a unified framework that can maintain consistent performance across both NSIs and SCIs. Addressing this issue requires approaches that are both perceptually aligned and adaptable to content variation without sacrificing computational efficiency.
\subsection{Motivation for a Robust Unified IQA Framework}
Given the limitations outlined above, a key research objective is to design an IQA framework that can generalize effectively across diverse image content types while remaining computationally feasible and perceptually relevant. To this end, we propose a hybrid strategy that leverages local structural complexity information through permutation entropy and employs a Random Forest regressor to model perceptual quality. The main contributions of our work are summarized as follows:
\begin{itemize} 
	\item A unified IQA framework that effectively handles both NSIs and SCIs by combining entropy-based features with machine learning. 
	\item A novel permutation entropy-based feature extraction scheme applied to gradient magnitude maps to capture perceptually relevant structures. 
	\item A full-reference IQA metric that balances interpretability, robustness, and computational simplicity. 
\end{itemize} 
These contributions are directly motivated by the shortcomings of content-specific IQA models and aim to bridge the generalization gap through content-agnostic feature modelling. While entropy-based techniques have shown promise, their stand-alone use has limitations in terms of adaptability and sensitivity to subtle distortions—challenges addressed through our integrated PEFRF design.
\subsection{Permutation Entropy for Perceptual Complexity Modelling}
\label{section:PEIQA}
Permutation Entropy (PE) \citep{bandt2002permutation} is a statistical measure used to quantify the complexity or disorder within a system based on the ordinal patterns of its values. Unlike traditional entropy measures that rely on histogram-based intensity distributions, PE captures the structural ordering of elements, making it especially effective in detecting fine-grained variations. This property has led to its successful application in fields such as biomedical signal analysis, EEG monitoring, and anomaly detection.\par
In the context of image quality assessment (IQA), PE offers a perceptually relevant means of capturing structural distortions by evaluating local ordering in pixel intensities. The normalized permutation entropy for a given sequence is computed as:
\begin{equation}
	PE_{nor}=-\frac{1}{(d-1)}\sum\limits_{c=1}^{d!}p_c log_2(p_c)
	\label{eq:PE}
\end{equation}
where $d$ is the permutation order (or embedding dimension), $p_c$ is the relative frequency of the $c$-th ordinal pattern, and $\tau$ is the delay parameter between successive elements in the sequence. A higher entropy value indicates greater spatial complexity and potential structural disorder in the input patch.\par
\begin{figure}[H]
	\centering
	\includegraphics[width=\columnwidth]{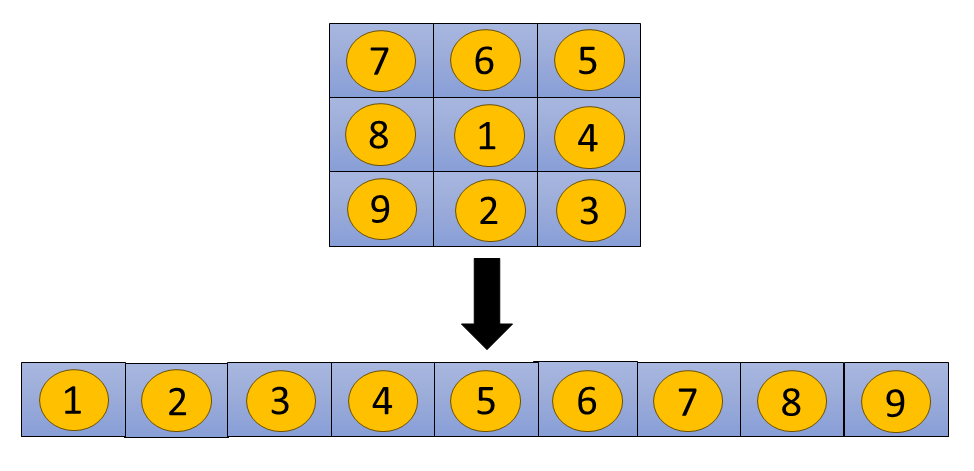}
	\caption{Conversion of a $3\times3$ image patch into a 1D sequence for permutation entropy computation.}
	\label{fig:PEImg}
\end{figure}

 
To adapt this concept to image data, each $3\times3$ pixel patch is treated as a one-dimensional sequence of nine intensity values. Using a sliding window approach, PE is computed for every patch across the image, resulting in a Permutation Entropy Map that characterizes local spatial complexity. This process is illustrated in Figure \ref{fig:PEImg}.
\begin{figure*}[htbp]
	\centering
	\includegraphics[width=\textwidth]{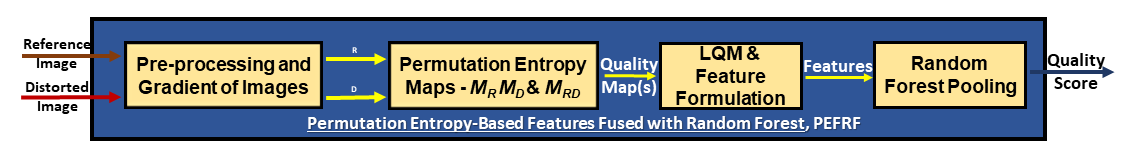}
	\caption{Block diagram of the proposed unified full-reference image quality assessment framework: permutation entropy-based features fused with random forest (PEFRF).}
	\label{fig:PEFRF}
\end{figure*}
In practice, we select a permutation order of $d=3$ and delay parameter $\tau=1$, which balances the granularity of structural representation with computational efficiency. These values have been shown in prior studies to produce stable and sensitive entropy estimates for small spatial regions. The resulting entropy maps act as compact yet informative representations of spatial complexity, reducing the dimensionality of the image data while preserving perceptually significant local structures.\par 
The entropy maps derived from the reference, distorted, and fused images serve as the input for subsequent stages of the PEFRF framework. By capturing localized changes in spatial regularity, these maps provide a perceptually meaningful signal for quality degradation, particularly in screen content images where structural integrity is critical. Their noise resilience and low computational cost also make them suitable for real-time or resource-constrained environments.
\subsection{Rationale Behind the PEFRF Framework Design}
The PEFRF framework is designed to address the core limitations of existing IQA methods: lack of content generalization, poor interpretability, and high computational demands. It combines the strengths of permutation entropy (PE) and ensemble learning to deliver a unified, content-agnostic approach to full-reference image quality assessment.\par 
Permutation entropy enables the capture of localized spatial complexity in images, providing a structural representation that is both perceptually meaningful and computationally lightweight. Unlike global image statistics or deep features that may require extensive training data, PE focuses on local ordering patterns, which align more closely with human visual perception—particularly for detecting distortions in both natural-scene and screen-content images.\par 
To model the relationship between the extracted entropy-based features and perceived image quality, the framework employs a Random Forest (RF) regressor. Random Forests are known for their robustness, generalization capability, and interpretability. They are also less prone to over-fitting compared to more complex neural network architectures, especially when training data is limited. This makes them a practical choice for scenarios where the model must perform reliably across varied content types and distortion conditions without requiring large-scale retraining.\par 
By integrating interpretable features derived from PE with a reliable learning-based regression model, PEFRF achieves a balance between accuracy, generalization, and computational efficiency. Its modular design allows it to be extended or adapted to specific content domains without re-engineering the entire framework. This makes it particularly well-suited for real-world deployment in applications where both structural sensitivity and content diversity are critical.
\section{Methodology}
The proposed PEFRF framework is a full-reference image quality assessment model designed to generalize across diverse content types by combining perceptually motivated entropy features with machine learning-based regression. The method operates in five sequential stages: (i) image preprocessing, (ii) gradient computation, (iii) local entropy map generation using permutation entropy, (iv) construction of a local quality map (LQM) and statistical feature extraction, and (v) final quality score prediction using a Random Forest regressor. The complete processing pipeline is illustrated in Figure \ref{fig:PEFRF}, which provides a high-level overview of the structural flow of the PEFRF framework.
\subsection{Preprocessing}
All images are first converted to greyscale to eliminate colour bias and reduce computational complexity, while preserving structural information relevant to human visual perception. Intensity normalization is then applied to standardize pixel value distributions across all images. These steps ensure consistency in gradient and entropy computations while maintaining the original perceptual characteristics of the dataset. As the same preprocessing is uniformly applied to both reference and distorted images, the benchmarking process remains fair and comparable across all methods.
\subsection{Gradient of Images}
Following the preprocessing stage, the PEFRF framework extracts gradient information from both the reference and distorted images. Gradients capture essential structural details and edge information, which are critical for human perception and effective image quality evaluation. By focusing on gradients, PEFRF emphasizes features that directly correlate with perceptual quality, such as texture, edges, and structural transitions within the image.\par
The gradient magnitudes are computed using the Sobel operator to enhance local variations:
\begin{equation}
	G = \sqrt{G_x^2 + G_y^2}
\end{equation}
where $G_x$ and $G_y$ are the horizontal and vertical gradients, respectively.\par 
Additionally, a third image is created by fusing the reference and distorted gradient images through pixel-wise averaging:
\begin{equation}
	G_{Fus} = \frac{1}{2}(G_{Ref} + G_{Dis})
\end{equation}
This fused image acts as a structural baseline for comparing mutual and divergent features between the reference and distorted inputs, enhancing the framework’s sensitivity to localized changes.
\subsection{Permutation Entropy Maps}
Permutation entropy maps are computed for three images: the reference image, the distorted image, and the fused reference-distorted image. These maps form a core component of the PEFRF framework by capturing local spatial complexity and randomness in pixel arrangements across the images. \par 
For each image, permutation entropy (PE) is computed using a sliding window of size $3\times3$ (sequence length 9), with an embedding dimension $d = 3$ and delay $\tau = 1$. These parameter values strike a balance between computational efficiency and the ability to capture localized structural patterns.\par 
The process begins by reshaping each $3\times3$ image patch into a 1D sequence of 9 values, as illustrated in Figure \ref{fig:PEImg}. Overlapping vectors of size 3 are then generated using a sliding $3\times1$ window on the sequence. Each vector is analysed to determine its ordinal pattern, and all $d!$ permutations are considered to compute a normalized probability distribution. \par 
Permutation entropy for each patch is calculated as the weighted sum of the probabilities of these ordinal patterns as in Equation \ref{eq:PE}. This process is repeated with one-pixel strides over the entire image for the reference, distorted, and fused gradient images, resulting in three corresponding PE maps: $m_R$, $m_D$, and $m_{RD}$. These maps effectively capture local spatial irregularities caused by distortion.\par 
The resulting entropy maps are used in the subsequent stage to generate the Local Quality Map (LQM), which enables entropy-based feature modelling for regression.
\subsection{Local Quality Map and Feature Extraction} 
The Local Quality Map (LQM) represents the degree of perceptual quality degradation across localized regions of an image. By capturing patch-wise quality levels, the LQM enables a fine-grained assessment of spatially distributed distortions, which is essential for accurate and context-aware image quality evaluation.\par
Using the permutation entropy maps $m_R$, $m_D$, and $m_{RD}$, corresponding to the reference, distorted, and fused gradient images respectively, the LQM is computed as defined in Equation \ref{eq:LQM}.\par 
\begin{equation}
	\begin{split}
		LQM = &\ \frac{2m_{R}m_{D}+\eta}{m_{R}^2+m_{D}^2+\eta} + \frac{2m_{D}m_{RD}+\eta}{m_{D}^2+m_{RD}^2+\eta} \\
		&+ \frac{2m_{R}m_{RD}+\eta}{m_{R}^2+m_{RD}^2+\eta}
	\end{split}
	\label{eq:LQM}
\end{equation}
Here, $eta$ is a small constant (set to 0.001) introduced to ensure numerical stability and prevent division errors when entropy values are close to zero. This value was chosen after extensive empirical evaluation to maintain a balance between sensitivity and robustness.\par
The resulting LQM reflects pixel-level distortion intensity using a structure-aware formulation. From this map, three statistical features are extracted to summarize the spatial distribution of local quality variation:
\begin{itemize}
	\item $f_1$: Median absolute deviation (MAD) of LQM values
	\item $f_2$: Standard deviation (SD) of LQM values
	\item $f_3$: Mean intensity (MEAN) of LQM values
\end{itemize}
These measures respectively capture the local variability and central tendency of perceived quality across the image. They are computed as in Equations \ref{eq:F1}, \ref{eq:F2}, and \ref{eq:F3}.
\begin{equation}
	f_1 = f_{MAD} = 1/N\sum_{i=1}^{N}\lvert LQM(i)-\overline{LQM}\rvert
	\label{eq:F1}
\end{equation}
\begin{equation}
	f_2= f_{SD} = \sqrt{1/N\sum_{i=1}^{N} \left(LQM(i)-\overline{LQM}\right)^2}
	\label{eq:F2}
\end{equation}
\begin{equation}
	f_3=f_{MEAN}= \frac{1}{N}\sum\limits_{i=1}^{N}LQM(i)
	\label{eq:F3}
\end{equation}
The final feature vector is:
\begin{equation}
	f_{FV}=\left[f_1,\, f_2,\, f_3\right]
	\label{eq:F}
\end{equation}
This vector serves as the input to the regression model in the next stage, enabling PEFRF to generate perceptually aligned quality estimates.
\subsection{Pooling Strategy - Regression using Random Forest} 
The pooling strategy in the PEFRF framework is designed to transform the localized quality assessments from the local quality map (LQM) into a single perceptual image quality score. While traditional pooling techniques—such as global averaging, weighted mean, or variance-based approaches—can summarize local quality variations, they often fail to capture the complex structural dependencies and spatial non-uniformities introduced by distortions. In many cases, these conventional methods risk suppressing or neutralizing important perceptual signals, especially in regions where distortion is localized or non-uniform.\par 
To address this limitation, PEFRF employs a regression-based pooling mechanism that uses a compact, three-dimensional feature vector derived from the LQM. This vector includes the mean, standard deviation, and median absolute deviation (MAD) of the LQM values. These features collectively encapsulate both the central tendency and variability of perceptual quality across the image, enabling the framework to account for spatial heterogeneity in distortion. \par 
The score computation is computer as in Equation \ref{eq:QS}.
\begin{equation}
	QS=\textit{RFR}(f_{FV}) \;\;\;\; \;\;\;\; \;\;\;\; where,\,0\leq QS\leq1 
	\label{eq:QS}
\end{equation} 
During the training phase, these LQM-derived feature vectors are paired with their corresponding ground-truth quality scores, typically Mean Opinion Scores (MOS) or Differential MOS (DMOS), to train a Random Forest regressor. This supervised learning step enables the model to learn the mapping between entropy-based structural features and subjective visual perception.\par 
\begin{strip}
	\begin{tcolorbox}[algobox, title=Algorithm 1: Permutation Entropy-Based Features Fused with Random Forest (PEFRF)]
		\textbf{Function:} $PEFRF(RI(x,y), DI(x,y), d, \tau, w_n)$ \\
		\textbf{Inputs:} $\underbrace{\text{Ref. Img.}}_{\text{GS}, RI(.) \in [0,1]} \rightarrow RI(x,y)$, 
		$\underbrace{\text{DI Img.}}_{\text{GS}, DI(.) \in [0,1]} \rightarrow DI(x,y)$, 
		Order $d=3$, Delay $\tau=1$, Window Size $w_n = 3 \times 3$ \\
		\textbf{Output:} Quality Score $Q_s \in [0,1]$\\
		\textbf{Initialization:}
		\begin{itemize}
			\item $T_r, T_c \gets \text{size}(RI(x,y))$
			\item $\eta = 0.001$
			\item Total permutation patterns: $\pi_j$, for order $d$ where $j = 1, \dots, d!$
		\end{itemize}
		\vspace{1mm}
		\textbf{Step 1:} \textbf{Edge Detection:} $R(x,y) \gets ED(RI(x,y))$, $D(x,y) \gets ED(DI(x,y))$ \\
		\textbf{Step 2:} $RD(x,y) \gets (R(x,y) + D(x,y)) / 2$\\
		\textbf{Step 3:} For $t = 1$ to $T_r$:
		\begin{itemize}
			\item For $i = 1$ to $T_c$, Step $\to w_n$:
			\begin{itemize}
				\item $Seq_R \gets R_{w_n}(t,:)$, $Seq_D \gets D_{w_n}(t,:)$, $Seq_{RD} \gets RD_{w_n}(t,:)$
				\item \textbf{Calculate Rank:} (for all sequences)
				\begin{itemize}
					\item $r_i, \dots, r_{i+n-1} \gets x_i, \dots, x_{i+n-1}$
					\item Where $r_i$ is the index of $x_i$ in ascending order
				\end{itemize}
				\item \textbf{Compare:} For each pattern $\pi_k$, where $k = 1, \dots, d!$ and $i = 1, \dots, i+n-1$
				\begin{itemize}
					\item If $\pi_k$ matches $r_i$, then $Z_k = Z_k + 1$
				\end{itemize}
				\item \textbf{Compute Probability:} $p_j' = z_j / \sum z_k$, for all $\pi_j$
				\item \textbf{Select:} $p_j' > 0$, for all $Seq_{(R,D,RD)}$
				\item \textbf{Compute:} 
				\[
				(mC_{R_i}, mC_{D_i}, mC_{RD_i}) = \sqrt{ \left( \frac{-1}{(d-1)} \sum_{j=1}^{d!} p_j' \log_2(p_j') \right)^2 }
				\]
			\end{itemize}
		\end{itemize}
		\vspace{1mm}
		\textbf{Step 4:} $mR_{R_t}, mR_{D_t}, mR_{RD_t} \gets mC_R, mC_D, mC_{RD}$ \\
		\textbf{Step 5:} $m_R, m_D, m_{RD} \gets mR_R, mR_D, mR_{RD}$ \\
		\textbf{Step 6:} Compute:
		\begin{itemize}
			\item Local Quality Map (LQM) \hfill (See Eq.~\ref{eq:LQM})
			\item Feature Vector $f_{FV}$ \hfill (See Eq.~\ref{eq:F})
			\item Final Score $Q_s \gets RFR(f_{FV})$ \hfill (See Eq.~\ref{eq:QS})
		\end{itemize}
	\end{tcolorbox}
\end{strip}
In the testing phase, feature vectors extracted from unseen distorted images are fed into the trained Random Forest regressor. The model outputs a predicted image quality score $QS$, which is normalized to lie within the range [0,1], where 0 indicates poor quality and 1 represents high perceptual fidelity. \par 
This regression-based pooling strategy not only preserves the integrity of local quality variations but also enhances interpretability and generalization. The use of Random Forest—a non-parametric ensemble method—offers several advantages: it avoids over-fitting, handles non-linear relationships between features and scores, and allows for feature importance analysis. Collectively, these properties make the PEFRF framework robust and adaptable for cross-content image quality assessment tasks.\par 
A complete overview of the PEFRF framework—from gradient-based preprocessing to entropy feature extraction and regression-based quality prediction—is outlined in Algorithm 1. This high-level pseudocode summarizes the full operational pipeline described throughout this section.
\section{Experimental Setup}
A comprehensive experimental setup was developed to validate the proposed PEFRF framework across diverse content types and distortion scenarios. The evaluation was conducted on 13 publicly available IQA benchmark datasets comprising over 21,000 images, and compared against more than 40 traditional and state-of-the-art IQA metrics. The assessment employed robust statistical validation methods—including Spearman (SRCC), Pearson (PLCC), Kendall (KRCC), Root Mean Square Error (RMSE), and F-test significance testing—to ensure reliable, reproducible, and perceptually aligned evaluation. This section details the datasets, implementation settings, and evaluation protocols used throughout the study.
\subsection{Datasets}
\begin{figure}[htbp]
	\centering
	\includegraphics[width=\columnwidth,keepaspectratio]{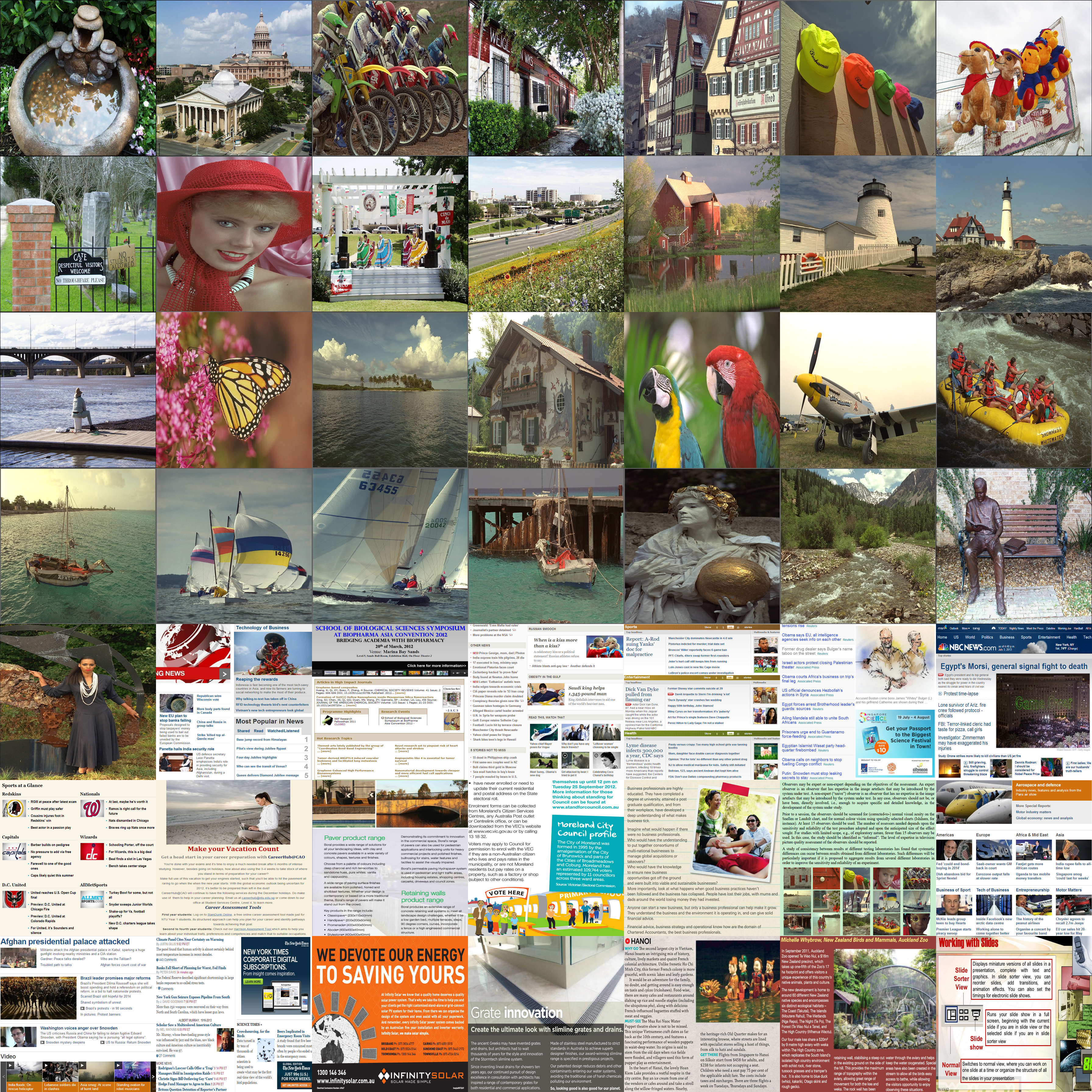}
	\caption{The figure illustrates reference natural-scene and screen-content images taken from LIVE and SIQAD datasets, respectively.}
	\label{figure:RDC}
\end{figure}

\begin{figure*}[htbp]
	\centering
	\includegraphics[width=0.75\textwidth,trim={3.5cm 0.5cm 3cm 0cm},clip]{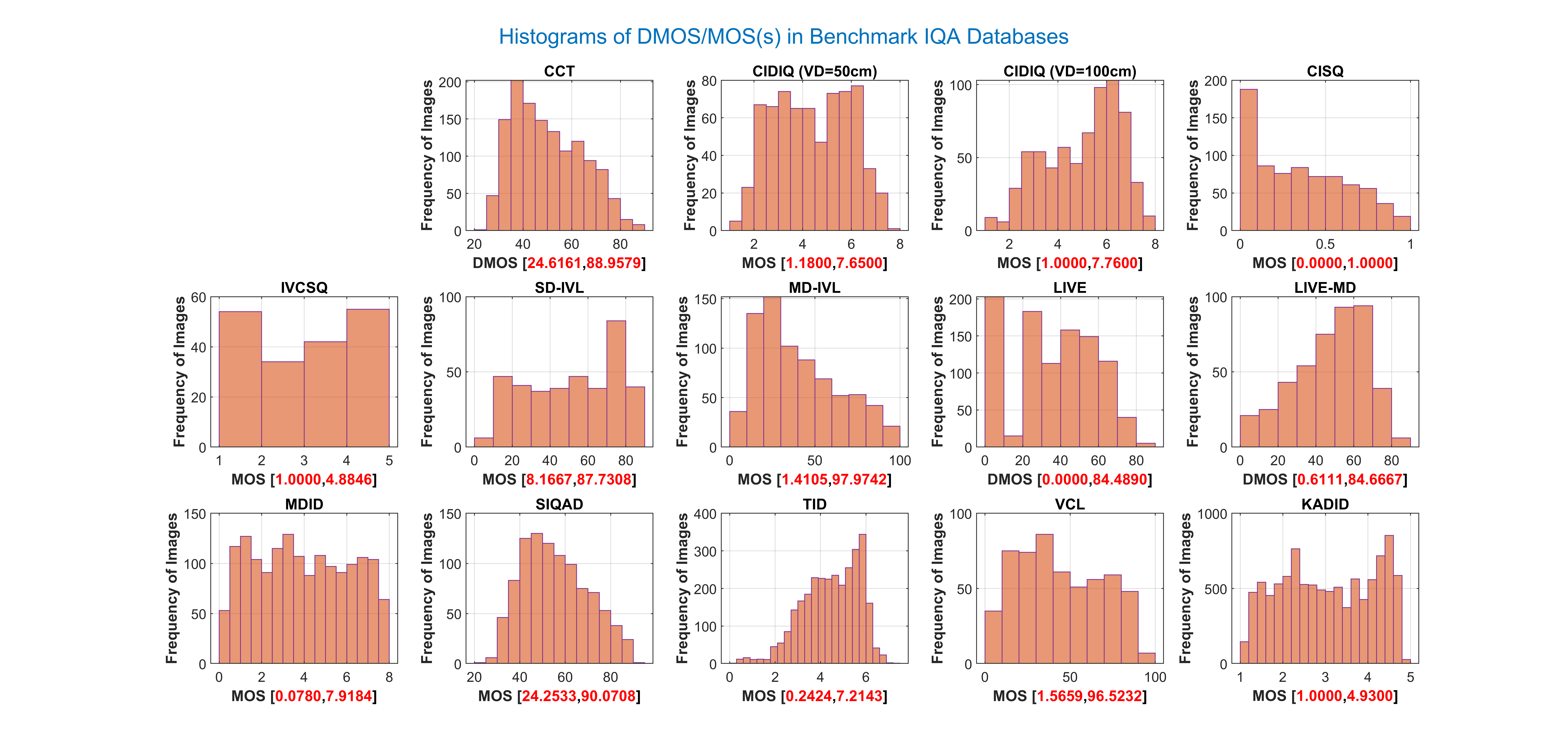}
	\caption{Histograms of the set of (difference) mean opinion scores (DMOS/MOS) of the benchmark datasets listed in Table. \ref{table:IQAdataset}.}
	\label{figure:MOSHist}
\end{figure*}

Robust benchmark datasets with annotated subjective scores, such as MOS and DMOS, were employed to evaluate PEFRF. Table \ref{table:IQAdataset} provides an overview of these datasets, showcasing their diversity in distortion types, image classes, and scoring ranges. Figures \ref{figure:RDC} and \ref{figure:MOSHist} illustrate sample images and the distribution of quality scores, emphasizing the datasets' comprehensiveness for both single- and multi-distortion scenarios.
\subsection{Implementation Details}
The proposed PEFRF framework was implemented as a robust full-reference image quality assessment framework designed for scalability and cross-content applicability. Key implementation details and computational configurations are as follows: 
\begin{enumerate}
	\item \textbf{Permutation Encoding:} Permutation entropy was calculated on overlapping image patches using an embedding dimension (\(d = 3\)) and a delay parameter (\(\tau = 1\)). These settings strike a balance between capturing meaningful pixel intensity fluctuations and minimizing computational overhead, ensuring computational efficiency.
	\item \textbf{RF Model Configuration:} The random forest (RF) model consisted of 200 decision trees, with a maximum tree depth of 20. The Gini impurity criterion was used for optimal splits, balancing model complexity and preventing over-fitting.
	\item \textbf{RF Hyper-parameter Tuning:} A grid search approach optimized the RF model parameters, such as the number of trees and maximum depth. This tuning enhanced prediction accuracy and ensured generalization across diverse datasets.
	\item \textbf{Score Mapping:} To align RF-predicted scores with human visual perception, a five-parameter logistic regression function was employed. This mapping ensured monotonicity with subjective evaluations, a critical aspect of perceptual image quality assessment.
	\item \textbf{Hardware:} Experiments were conducted on a workstation equipped with an Intel Core i7 processor, 16 GB RAM, and an NVIDIA GPU. This setup provided sufficient computational power for feature extraction, model training, and evaluation.
\end{enumerate}
\begin{table*}[t]
	\centering
	\caption{Details of benchmark dataset(s) used for performance evaluation of the proposed PEFRF FR-IQA framework.}
	\label{table:IQAdataset}
	\setlength{\tabcolsep}{2pt}
	\renewcommand{\arraystretch}{1}
	\footnotesize
	\begin{tabularx}{0.95\textwidth}{C{0.5cm}|p{6cm}|C{0.75cm}|C{0.75cm}|C{0.75cm}|C{0.75cm}|C{1.5cm}|C{1.9cm}|C{2.1cm}|C{0.4cm}}
		\hline
		\textbf{SN} &  \multicolumn{1}{c|}{\textbf{Dataset(s)}} & \textbf{Year} & \textbf{NRI} & \textbf{NDI} & \textbf{NDT} & \textbf{DT(s)} & \textbf{Image Class} & \textbf{Score Range} & \textbf{DT} \\
		\hline \hline
		1 & CCT \citep{8000398} & 2005 & 74 & 1320 & 2 & Artificial & NSI, SCI, CGI & DMOS [0,100] & SDT \\
		2 & CIDIQ \citep{liu2014cid} & 2013 & 23 & 980 & 6 & Artificial & NSI & MOS [0,9] & SDT \\
		3 & CSIQ \citep{larson2010most} & 2010 & 30 & 866 & 6 & Artificial & NSI & MOS [0,1] & SDT \\
		4 & IVCSQ \citep{site2subjective} & 2005 & 10 & 235 & 5 & Artificial & NSI & MOS [1,5] & SDT \\
		5 & SD-IVL \citep{corchs2014no-reference} & 2014 & 20 & 380 & 2 & Artificial & NSI & MOS [0,100] & SDT \\
		6 & MD-IVL \citep{corchs2017multidistortion} & 2017 & 10 & 750 & 2 & Artificial & NSI & MOS [0,100] & MDT \\
		7 & LIVE \citep{sheikh2006statistical,wang2004image} & 2006 & 29 & 779 & 6 & Artificial & NSI & DMOS [0,100] & SDT \\
		8 & LIVE-MD \citep{jayaraman2012objective} & 2012 & 15 & 450 & 5 & Artificial & NSI & DMOS [0,100] & MDT \\
		9 & MDID \citep{sun2017mdid} & 2017 & 20 & 1600 & 5 & Artificial & NSI & MOS [0,8] & MDT \\
		10 & SIQAD \citep{yang2015perceptual} & 2015 & 20 & 980 & 7 & Artificial & SCI & MOS [0,100] & SDT \\
		11 & TID 2013 \citep{ponomarenko2015image} & 2013 & 25 & 3000 & 24 & Artificial & NSI & MOS [0,9] & SDT \\
		12 & VCL \citep{zaric2012vcl} & 2012 & 23 & 552 & 4 & Artificial & NSI & MOS [0,100] & SDT \\
		13 & KADID \citep{kadid10k} & 2019 & 81 & 10125 & 25 & Artificial & NSI & MOS [1,5] & SDT \\
		\hline
	\end{tabularx}
	\vspace{1ex}
	\begin{minipage}{0.8\linewidth}
		\footnotesize
		\begin{tabular}{@{}p{0.31\linewidth} p{0.31\linewidth} p{0.31\linewidth}@{}}
			\textbf{NRI(s)} – No. of Reference Images & \textbf{NDI(s)} – No. of Distorted Images & \textbf{NDT(s)} – No. of Distortion Types \\
			\textbf{DT(s)} – Dataset Type & \textbf{NSI} – Natural-Scene Images  & \textbf{SCI} – Screen-Content Images  \\
			\textbf{CGI} – Computer-Graphic Images & \textbf{SDT} – Single Distortion Type & \textbf{MDT} – Multiple Distortion Type \\
		\end{tabular}
	\end{minipage}
\end{table*}
\subsection{Cross-content Applicability}
The PEFRF framework demonstrated strong generalization across diverse content types, including natural-scene and screen-content images. By leveraging gradient maps and permutation entropy, the framework effectively captured subtle structural distortions, showcasing its versatility for various image quality assessment tasks.
\subsection{Training and Validation}
To ensure robustness and reliability, the following procedures were employed:
\begin{enumerate}
	\item \textbf{Cross-Validation:}A five-fold cross-validation scheme was adopted to evaluate the performance of the Random Forest Regressor. This approach ensured consistent performance across different subsets of training data, providing a reliable mechanism to validate hyper-parameter configurations and reduce the risk of over-fitting.
	\item \textbf{Out-of-Bag (OOB) Error Estimates:} During training, Out-of-Bag (OOB) error estimates, an intrinsic feature of the Random Forest algorithm, were monitored. OOB errors provided an unbiased assessment of the model's performance and served as an additional metric to guide hyper-parameter tuning.
\end{enumerate}
\subsection{Performance Comparison}
This section outlines the metrics used to compare the proposed PEFRF framework in the context of image quality assessment (IQA). The evaluation involved a broad spectrum of conventional and state-of-the-art (SOTA) IQA metrics, ensuring a comprehensive assessment of PEFRF's effectiveness. The comparison was conducted across both full-reference and no-reference IQA metrics, highlighting the robustness and generalizability of PEFRF.\par 
The conventional IQA metrics included foundational methods such as mean squared error (MSE), peak signal-to-noise ratio (PSNR), information-weighted mean square error (IW-MSE) (\citep{wang2011information}), and information-weighted peak signal-to-noise ratio (IW-PSNR) (\citep{wang2011information}). Additionally, advanced metrics like universal image quality index (UIQI) (\citep{wang2001demo}), structural similarity index (SSIM) (\citep{wang2004image}), multi-scale structural similarity (MS-SSIM) (\citep{wang2003multiscale}), gradient magnitude similarity deviation (GMSD) (\citep{Xue2014}), and Haar Wavelet-based perceptual similarity index (HaarPSI) (\citep{reisenhofer2018haar}) were included to ensure coverage of diverse evaluation criteria.\par 
Furthermore, the comparison included SOTA metrics such as perceptual similarity (PSIM) (\citep{gu2017fast}), perceptual image-error assessment through pairwise preference (PieAPP) (\citep{prashnani2018pieapp}), and deep image structure and texture similarity (DISTS) (\citep{ding2020image}). No-reference metrics like blind image quality measure of enhanced images (BIQME) (\citep{gu2017learning}), natural image quality evaluator (NIQE) (\citep{mittal2012making}), and blind/reference-less image spatial quality evaluator (BRISQE) (\citep{mittal2012no}) were also considered to evaluate PEFRF’s adaptability in scenarios where no reference image is available.\par 
In total, the proposed PEFRF framework was compared against over forty IQA metrics, spanning conventional and SOTA methods. This broad spectrum of comparison metrics allowed for a detailed evaluation of PEFRF's strengths and weaknesses, ensuring a thorough assessment of its performance. The inclusion of diverse metrics ensured that PEFRF’s performance was tested across a wide range of criteria, including perceptual fidelity, structural similarity, and robustness to various types of distortions.
\subsection{Performance Evaluation Criteria}
The evaluation of the proposed PEFRF framework is based on its ability to predict perceptual image quality scores that correlate strongly with subjective scores from benchmark IQA datasets. Subjective scores, such as mean opinion scores (MOS) or differential mean opinion scores (DMOS), are derived from human observer studies and serve as ground truth for comparison. To assess the effectiveness of PEFRF, four widely used performance evaluation criteria were utilized: Pearson linear correlation coefficient (PLCC), Spearman rank-order correlation coefficient (SRCC), Kendall rank-order correlation coefficient (KRCC), and root mean square error (RMSE). \par 
PLCC evaluates the prediction accuracy, while SRCC and KRCC measure the monotonicity between predicted and subjective scores. RMSE quantifies the prediction consistency by measuring the deviation between predicted and subjective scores. Higher correlation values (closer to 1) for PLCC, SRCC, and KRCC indicate better alignment with subjective assessments, while a lower RMSE reflects greater prediction consistency. To ensure the robustness of the PEFRF framework, a five-parameter logistic regression function, as given in Equation \ref{eq:LR}, was used to map predicted objective scores to subjective scores:
\begin{equation}
	\label{eq:LR}
	Q_p = q_1\left(\frac{1}{2} - \frac{1}{1 + e^{q_2(Q - q_3)}}\right) + q_4Q + q_5
\end{equation}
Here, $Q$ and $Q_p$ are the original and mapped IQA scores, respectively, while $q_i (i=1,2,3,4,5)$ are coefficients estimated through curve fitting.
After regression, the performance metrics are computed as follows: \vspace{10pt} 
\subsubsection{Spearman Rank-Order Correlation Coefficient (SRCC)} Measures monotonicity between predicted and subjective scores. Mathematically, it is defined as:
\begin{equation}
	\label{eq:SRCC}
	SRCC(Q,S) = 1 - \frac{6\sum_{i=1}^{n}d_i^2}{n(n^2-1)}
\end{equation}
where $d_i$ is the rank difference for each pair of samples, and $n$ is the total number of samples.
\subsubsection{Kendall Rank-Order Correlation Coefficient (KRCC)} Evaluates the prediction monotonicity using concordant ($n_c$) and discordant ($n_d$) pairs:
\begin{equation}
	\label{eq:KRCC}
	KRCC(Q,S) = \frac{n_c - n_d}{0.5n(n-1)}
\end{equation}
\subsubsection{Pearson Linear Correlation Coefficient (PLCC)} Measures the linear correlation between the mapped IQA scores $Q_p$ and subjective scores $S$:
\begin{equation}
	\label{eq:PLCC}
	PLCC(Q_p,S) = \frac{\overline{Q}_p^T\overline{S}}{\sqrt{\overline{Q}_p^T\overline{Q}_p \cdot \overline{S}^T\overline{S}}}
\end{equation}
where $\overline{Q}_p$ and $\overline{S}$ are zero-mean objective and subjective scores, respectively.
\subsubsection{Root Mean Square Error (RMSE)} Reflects the deviation between mapped objective scores and subjective scores:
\begin{equation}
	\label{eq:RMSE}
	RMSE(Q_p,S) = \sqrt{\frac{(Q_p - S)^T(Q_p - S)}{n}}
\end{equation}
These performance metrics provide a comprehensive evaluation of the PEFRF framework’s ability to predict image quality accurately and consistently while maintaining coherence with human perceptual judgments.
\section{Results and Discussion}
This section evaluates the proposed PEFRF framework through a comprehensive analysis of its performance on benchmark IQA datasets. The assessment employs both quantitative and qualitative methods to measure its correlation with subjective quality scores. Quantitative evaluation involves widely accepted criteria, including the Pearson linear correlation coefficient (PLCC), Spearman rank-order correlation coefficient (SRCC), Kendall’s rank-order correlation coefficient (KRCC), and root mean square error (RMSE). Qualitative analysis explores specific cases to demonstrate the metric’s behaviour across diverse image distortions, with the F-test applied to assess the statistical significance of performance differences between PEFRF and competing IQA methods. The F-test is suitable in this context as it evaluates the equality of variances across prediction errors, helping to determine whether the observed differences in metric performance are statistically meaningful. Together, these analyses confirm the robustness and generalizability of the PEFRF framework for cross-content image quality assessment.
\subsection{Quantitative Analysis}
The proposed PEFRF framework was rigorously evaluated across multiple datasets representative of cross-content applications. Performance metrics such as PLCC, SRCC, KRCC, and RMSE were computed to measure predictive accuracy, ranking consistency, and error minimization in quality predictions. To ensure the reliability of observed improvements, an F-test was conducted to verify that the performance gains are statistically significant and not due to random fluctuations. \par 
Tables \ref{table:IQAComp1} and \ref{table:IQAComp2} summarize the comparative performance of the proposed PEFRF framework against state-of-the-art IQA algorithms across 13 benchmark datasets. These tables highlight the PEFRF framework's ability to consistently achieve high prediction accuracy and ranking consistency. Additionally, Table \ref{table:IQAComp3} provides a detailed analysis of PEFRF’s performance across major distortion types, showcasing its reliability in handling diverse image quality variations.
\begin{table*}[htbp]
	\centering
	\caption{Results of performance comparison between proposed PEFRF framework with SOTA IQA metrics on benchmark datasets based on SRCC and PLCC values. \textbf{Note:} Top-five performing models are indicated with bold and coloured fonts, i.e., \color{red}\textbf{$1^{ST}$}\color{black}, \color{blue}\textbf{$2^{ND}$}\color{black}, \color{blue}\textbf{$3^{RD}$}\color{black}, \color{blue}\textbf{$4^{TH}$}\color{black}, and \color{blue}\textbf{$5^{TH}$}\color{black}.}
	\label{table:IQAComp1}
	\setlength\tabcolsep{0.2pt}
	\setstretch{1.5}
	\begin{adjustbox}{width=\textwidth}
\begin{tabular}{
				S[table-format=1.4]
				S[table-format=1.4] S[table-format=1.4]
				S[table-format=1.4] S[table-format=1.4]
				S[table-format=1.4] S[table-format=1.4]
				S[table-format=1.4] S[table-format=1.4]
				S[table-format=1.4] S[table-format=1.4]
				S[table-format=1.4] S[table-format=1.4]
				S[table-format=1.4] S[table-format=1.4]
				S[table-format=1.4] S[table-format=1.4]
				S[table-format=1.4] S[table-format=1.4]
				S[table-format=1.4] S[table-format=1.4]
				S[table-format=1.4] S[table-format=1.4]
				S[table-format=1.4] S[table-format=1.4]
				S[table-format=1.4] S[table-format=1.4]
			}
			\toprule
			\scriptsize\textbf{Metric}&\multicolumn{2}{c}{\scriptsize\textbf{CCT}}& \multicolumn{2}{c}{\scriptsize\textbf{CIDIQ}}& \multicolumn{2}{c}{\scriptsize\textbf{CSIQ}}& \multicolumn{2}{c}{\scriptsize\textbf{IVC}}& \multicolumn{2}{c}{\scriptsize\textbf{SD-IVL}}& \multicolumn{2}{c}{\scriptsize\textbf{MD-IVL}}& \multicolumn{2}{c}{\scriptsize\textbf{LIVE}}& \multicolumn{2}{c}{\scriptsize\textbf{LIVE-MD}}& \multicolumn{2}{c}{\scriptsize\textbf{MDID}}& \multicolumn{2}{c}{\scriptsize\textbf{SIQAD}}& \multicolumn{2}{c}{\scriptsize\textbf{TID}}& \multicolumn{2}{c}{\scriptsize\textbf{VCL}}& \multicolumn{2}{c}{\scriptsize\textbf{KADID}}\\
			\midrule
			&{\scriptsize\textbf{SRCC}}&{\scriptsize\textbf{PLCC}}&{\scriptsize\textbf{SRCC}}&{\scriptsize\textbf{PLCC}}&{\scriptsize\textbf{SRCC}}&{\scriptsize\textbf{PLCC}}&{\scriptsize\textbf{SRCC}}&{\scriptsize\textbf{PLCC}}&{\scriptsize\textbf{SRCC}}&{\scriptsize\textbf{PLCC}}&{\scriptsize\textbf{SRCC}}&{\scriptsize\textbf{PLCC}}&{\scriptsize\textbf{SRCC}}&{\scriptsize\textbf{PLCC}}&{\scriptsize\textbf{SRCC}}&{\scriptsize\textbf{PLCC}}&{\scriptsize\textbf{SRCC}}&{\scriptsize\textbf{PLCC}}&{\scriptsize\textbf{SRCC}}&{\scriptsize\textbf{PLCC}}&{\scriptsize\textbf{SRCC}}&{\scriptsize\textbf{PLCC}}&{\scriptsize\textbf{SRCC}}&{\scriptsize\textbf{PLCC}}&{\scriptsize\textbf{SRCC}}&{\scriptsize\textbf{PLCC}}\\
\hline
			\bottomrule
	\end{tabular}}
\end{adjustbox}
\vspace{1mm}
\noindent
\begin{tabularx}{\linewidth}{@{}X p{0.35\linewidth}@{}}
	\crule[uclagold]{6pt}{6pt} - Full-Reference IQA Metric
	& \crule[cyan(process)]{6pt}{6pt} - No-Reference IQA Metric \\
	\crule[bubblegum]{6pt}{6pt} - Proposed FR-IQA Metric
	& \\
\end{tabularx}
\end{table*}
\begin{table*}[htbp]
	\centering
	\caption{Results of performance comparison between proposed PEFRF framework with SOTA IQA metrics on benchmark datasets based on KRCC and RMSE values. \textbf{Note:} Top-five performing models are indicated with bold and coloured fonts, i.e., \color{red}\textbf{$1^{ST}$}\color{black}, \color{blue}\textbf{$2^{ND}$}\color{black}, \color{blue}\textbf{$3^{RD}$}\color{black}, \color{blue}\textbf{$4^{TH}$}\color{black}, and \color{blue}\textbf{$5^{TH}$}\color{black}.}
	\label{table:IQAComp2}
	\setlength\tabcolsep{0.2pt}
	\setstretch{1.5}
	\begin{adjustbox}{width=\textwidth}
\begin{tabular}{
				l
				S[table-format=1.4] S[table-format=1.4]
				S[table-format=1.4] S[table-format=1.4]
				S[table-format=1.4] S[table-format=1.4]
				S[table-format=1.4] S[table-format=1.4]
				S[table-format=1.4] S[table-format=1.4]
				S[table-format=1.4] S[table-format=1.4]
				S[table-format=1.4] S[table-format=1.4]
				S[table-format=1.4] S[table-format=1.4]
				S[table-format=1.4] S[table-format=1.4]
				S[table-format=1.4] S[table-format=1.4]
				S[table-format=1.4] S[table-format=1.4]
				S[table-format=1.4] S[table-format=1.4]
				S[table-format=1.4] S[table-format=1.4]
			}
			\toprule
			\scriptsize\textbf{Metric}&\multicolumn{2}{c}{\scriptsize\textbf{CCT}}& \multicolumn{2}{c}{\scriptsize\textbf{CIDIQ}}& \multicolumn{2}{c}{\scriptsize\textbf{CSIQ}}& \multicolumn{2}{c}{\scriptsize\textbf{IVC}}& \multicolumn{2}{c}{\scriptsize\textbf{SD-IVL}}& \multicolumn{2}{c}{\scriptsize\textbf{MD-IVL}}& \multicolumn{2}{c}{\scriptsize\textbf{LIVE}}& \multicolumn{2}{c}{\scriptsize\textbf{LIVE-MD}}& \multicolumn{2}{c}{\scriptsize\textbf{MDID}}& \multicolumn{2}{c}{\scriptsize\textbf{SIQAD}}& \multicolumn{2}{c}{\scriptsize\textbf{TID}}& \multicolumn{2}{c}{\scriptsize\textbf{VCL}}& \multicolumn{2}{c}{\scriptsize\textbf{KADID}}\\
			\midrule
			&{\scriptsize\textbf{KRCC}}&{\scriptsize\textbf{RMSE}}&{\scriptsize\textbf{KRCC}}&{\scriptsize\textbf{RMSE}}&{\scriptsize\textbf{KRCC}}&{\scriptsize\textbf{RMSE}}&{\scriptsize\textbf{KRCC}}&{\scriptsize\textbf{RMSE}}&{\scriptsize\textbf{KRCC}}&{\scriptsize\textbf{RMSE}}&{\scriptsize\textbf{KRCC}}&{\scriptsize\textbf{RMSE}}&{\scriptsize\textbf{KRCC}}&{\scriptsize\textbf{RMSE}}&{\scriptsize\textbf{KRCC}}&{\scriptsize\textbf{RMSE}}&{\scriptsize\textbf{KRCC}}&{\scriptsize\textbf{RMSE}}&{\scriptsize\textbf{KRCC}}&{\scriptsize\textbf{RMSE}}&{\scriptsize\textbf{KRCC}}&{\scriptsize\textbf{RMSE}}&{\scriptsize\textbf{KRCC}}&{\scriptsize\textbf{RMSE}}&{\scriptsize\textbf{KRCC}}&{\scriptsize\textbf{RMSE}}\\
\hline
			
			\bottomrule
	\end{tabular}}
	\vspace{1mm}
	\noindent
	\begin{tabularx}{\linewidth}{@{}X p{0.35\linewidth}@{}}
		\hspace{2em}\crule[uclagold]{6pt}{6pt} - Full-Reference IQA Metric
		& \crule[cyan(process)]{6pt}{6pt} - No-Reference IQA Metric \\
		\hspace{2em}\crule[bubblegum]{6pt}{6pt} - Proposed FR-IQA Metric
		& \\
	\end{tabularx}
	\end{adjustbox}
\end{table*}
\begin{table*}[htbp]
	\centering
	\caption{Results of distortion-wise performance comparison of the proposed PEFRF framework with SOTA IQA metrics on benchmark datasets based on SRCC, PLCC, KRCC, and RMSE values. \textbf{Note:} Top-five performing metrics are indicated with bold and coloured fonts, i.e., \color{red}\textbf{$1^{ST}$}\color{black}, \color{blue}\textbf{$2^{ND}$}\color{black}, \color{blue}\textbf{$3^{RD}$}\color{black}, \color{blue}\textbf{$4^{TH}$}\color{black}, and \color{blue}\textbf{$5^{TH}$}\color{black}.}
	\label{table:IQAComp3}
	\setlength\tabcolsep{0.3pt}
	\setstretch{1}
		\begin{adjustbox}{width=\textwidth}
\begin{tabular}{
				l
				S[table-format=1.4] S[table-format=1.4]
				S[table-format=1.4] S[table-format=1.4]
				S[table-format=1.4] S[table-format=1.4]
				S[table-format=1.4] S[table-format=1.4]
				S[table-format=1.4] S[table-format=1.4]
				S[table-format=1.4] S[table-format=1.4]
				S[table-format=1.4] S[table-format=1.4]
				S[table-format=1.4] S[table-format=1.4]
			}
			\toprule
			\multicolumn{17}{c}{\footnotesize \textbf{Distortion-wise Performance Evaluation of IQA Metrics}}  \\\hline
			\multirow{3}{*}{\scriptsize\textbf{Metrics}} & \multicolumn{16}{c}{\footnotesize{\textbf{Performance Evaluation Criteria}}}  \\ \cline{2-17}
			
			& \multicolumn{4}{c|}{\scriptsize\textbf{SRCC}} & \multicolumn{4}{c|}{\scriptsize\textbf{PLCC}} & \multicolumn{4}{c|}{\scriptsize\textbf{KRCC}} & \multicolumn{4}{c}{\scriptsize\textbf{RMSE}}\\ \cline{2-17}
			
			& \multicolumn{1}{c|}{\scriptsize{GN}} & \multicolumn{1}{c|}{\scriptsize{GB}} & \multicolumn{1}{c|}{\scriptsize{JPEG}} & \multicolumn{1}{c|}{\scriptsize{J2K}} & \multicolumn{1}{c|}{\scriptsize{GN}} & \multicolumn{1}{c|}{\scriptsize{GB}} & \multicolumn{1}{c|}{\scriptsize{JPEG}} & \multicolumn{1}{c|}{\scriptsize{J2K}} & \multicolumn{1}{c|}{\scriptsize{GN}} & \multicolumn{1}{c|}{\scriptsize{GB}} & \multicolumn{1}{c|}{\scriptsize{JPEG}} & \multicolumn{1}{c|}{\scriptsize{J2K}} & \multicolumn{1}{c|}{\scriptsize{GN}} & \multicolumn{1}{c|}{\scriptsize {GB}} & \multicolumn{1}{c|}{\scriptsize{JPEG}} & \multicolumn{1}{c}{\scriptsize{J2K}}  \\
\hline
			
			\bottomrule
	\end{tabular}
		\end{adjustbox}
	\vspace{1mm}
	\noindent
		\begin{tabularx}{\linewidth}{@{}X p{0.35\linewidth}@{}}
			\hspace{2em}\crule[uclagold]{6pt}{6pt} - Full-Reference IQA Metric
			& \crule[cyan(process)]{6pt}{6pt} - No-Reference IQA Metric \\
			\hspace{2em}\crule[bubblegum]{6pt}{6pt} - Proposed FR-IQA Metric
			& \\
		\end{tabularx}
\end{table*}
\setlength\tabcolsep{2pt}
\setstretch{0.9}
\begin{table*}[htbp]
	\centering
	\caption{Results of performance comparison between the proposed PEFRF FR-IQA framework and SOTA IQA metrics, based on statistical significance test (F-Test) on the CIDIQ dataset.}
	\label{table:ST11}
	\begin{tabular}{|>{\arraybackslash\centering\tiny}p{0.2cm}
			|>{\tiny}p{2cm}
			|>{\arraybackslash\centering\tiny}p{0.2cm}
			|>{\arraybackslash\centering\tiny}p{0.2cm}
			|>{\arraybackslash\centering\tiny}p{0.2cm}
			|>{\arraybackslash\centering\tiny}p{0.2cm}
			|>{\arraybackslash\centering\tiny}p{0.2cm}
			|>{\arraybackslash\centering\tiny}p{0.2cm}
			|>{\arraybackslash\centering\tiny}p{0.2cm}
			|>{\arraybackslash\centering\tiny}p{0.2cm}
			|>{\arraybackslash\centering\tiny}p{0.2cm}
			|>{\arraybackslash\centering\tiny}p{0.2cm}
			|>{\arraybackslash\centering\tiny}p{0.2cm}
			|>{\arraybackslash\centering\tiny}p{0.2cm}
			|>{\arraybackslash\centering\tiny}p{0.2cm}
			|>{\arraybackslash\centering\tiny}p{0.2cm}
			|>{\arraybackslash\centering\tiny}p{0.2cm}
			|>{\arraybackslash\centering\tiny}p{0.2cm}
			|>{\arraybackslash\centering\tiny}p{0.2cm}
			|>{\arraybackslash\centering\tiny}p{0.2cm}
			|>{\arraybackslash\centering\tiny}p{0.2cm}
			|>{\arraybackslash\centering\tiny}p{0.2cm}
			|>{\arraybackslash\centering\tiny}p{0.2cm}
			|>{\arraybackslash\centering\tiny}p{0.2cm}
			|>{\arraybackslash\centering\tiny}p{0.2cm}
			|>{\arraybackslash\centering\tiny}p{0.2cm}
			|>{\arraybackslash\centering\tiny}p{0.2cm}
			|>{\arraybackslash\centering\tiny}p{0.2cm}
			|>{\arraybackslash\centering\tiny}p{0.2cm}
			|>{\arraybackslash\centering\tiny}p{0.2cm}
			|>{\arraybackslash\centering\tiny}p{0.2cm}
			|>{\arraybackslash\centering\tiny}p{0.2cm}
			|>{\arraybackslash\centering\tiny}p{0.2cm}
			|>{\arraybackslash\centering\tiny}p{0.2cm}
			|>{\arraybackslash\centering\tiny}p{0.2cm}
			|>{\arraybackslash\centering\tiny}p{0.2cm}
			|>{\arraybackslash\centering\tiny}p{0.2cm}
			|>{\arraybackslash\centering\tiny}p{0.2cm}
			|>{\arraybackslash\centering\tiny}p{0.2cm}
			|>{\arraybackslash\centering\tiny}p{0.2cm}
			|>{\arraybackslash\centering\tiny}p{0.2cm}
			|>{\arraybackslash\centering\tiny}p{0.2cm}
			|>{\arraybackslash\centering\tiny}p{0.2cm}
			|>{\arraybackslash\centering\tiny}p{0.2cm}
			|>{\arraybackslash\centering\tiny}p{0.2cm}
			|>{\arraybackslash\centering\tiny}p{0.2cm}
			|>{\arraybackslash\centering\tiny}p{0.2cm}
			|>{\arraybackslash\centering\tiny}p{0.2cm}
			|>{\arraybackslash\centering\tiny}p{0.2mm}|
		}
		\toprule
		\hline
		\multicolumn{46}{c}{\small\bfseries Statistical Significance Test - CIDIQ dataset (F-Critical: \color{red}1.0619\color{black})}  \\
\hline
		
		\bottomrule
	\end{tabular}
	\vspace{1mm}
	\noindent
	\begin{tabularx}{\linewidth}{@{}X p{0.35\linewidth}@{}}
		\hspace{2em}\crule[black]{6pt}{6pt} - Statistically Significant
		& \crule[unmyellow]{6pt}{6pt} - Statistically Insignificant\\
		\hspace{2em}\crule[lime(colorwheel)]{6pt}{6pt} - Statistically Indistinguishable
		&\crule[uclagold]{6pt}{6pt} - Top-5 Performing Metrics\\
		\hspace{2em}\crule[lavender(floral)]{6pt}{6pt} - Not Applicable&\\
	\end{tabularx}
\end{table*}
\setlength\tabcolsep{2pt}
\setstretch{0.9}
\begin{table*}[htbp]
	\centering
	\caption{Results of performance comparison between the proposed PEFRF FR-IQA framework and SOTA IQA metrics, based on statistical significance test (F-Test) on the SIQAD dataset.}
	\label{table:ST10}
	\begin{tabular}{|>{\arraybackslash\centering\tiny}p{0.2cm}
			|>{\tiny}p{2cm}
			|>{\arraybackslash\centering\tiny}p{0.2cm}
			|>{\arraybackslash\centering\tiny}p{0.2cm}
			|>{\arraybackslash\centering\tiny}p{0.2cm}
			|>{\arraybackslash\centering\tiny}p{0.2cm}
			|>{\arraybackslash\centering\tiny}p{0.2cm}
			|>{\arraybackslash\centering\tiny}p{0.2cm}
			|>{\arraybackslash\centering\tiny}p{0.2cm}
			|>{\arraybackslash\centering\tiny}p{0.2cm}
			|>{\arraybackslash\centering\tiny}p{0.2cm}
			|>{\arraybackslash\centering\tiny}p{0.2cm}
			|>{\arraybackslash\centering\tiny}p{0.2cm}
			|>{\arraybackslash\centering\tiny}p{0.2cm}
			|>{\arraybackslash\centering\tiny}p{0.2cm}
			|>{\arraybackslash\centering\tiny}p{0.2cm}
			|>{\arraybackslash\centering\tiny}p{0.2cm}
			|>{\arraybackslash\centering\tiny}p{0.2cm}
			|>{\arraybackslash\centering\tiny}p{0.2cm}
			|>{\arraybackslash\centering\tiny}p{0.2cm}
			|>{\arraybackslash\centering\tiny}p{0.2cm}
			|>{\arraybackslash\centering\tiny}p{0.2cm}
			|>{\arraybackslash\centering\tiny}p{0.2cm}
			|>{\arraybackslash\centering\tiny}p{0.2cm}
			|>{\arraybackslash\centering\tiny}p{0.2cm}
			|>{\arraybackslash\centering\tiny}p{0.2cm}
			|>{\arraybackslash\centering\tiny}p{0.2cm}
			|>{\arraybackslash\centering\tiny}p{0.2cm}
			|>{\arraybackslash\centering\tiny}p{0.2cm}
			|>{\arraybackslash\centering\tiny}p{0.2cm}
			|>{\arraybackslash\centering\tiny}p{0.2cm}
			|>{\arraybackslash\centering\tiny}p{0.2cm}
			|>{\arraybackslash\centering\tiny}p{0.2cm}
			|>{\arraybackslash\centering\tiny}p{0.2cm}
			|>{\arraybackslash\centering\tiny}p{0.2cm}
			|>{\arraybackslash\centering\tiny}p{0.2cm}
			|>{\arraybackslash\centering\tiny}p{0.2cm}
			|>{\arraybackslash\centering\tiny}p{0.2cm}
			|>{\arraybackslash\centering\tiny}p{0.2cm}
			|>{\arraybackslash\centering\tiny}p{0.2cm}
			|>{\arraybackslash\centering\tiny}p{0.2cm}
			|>{\arraybackslash\centering\tiny}p{0.2cm}
			|>{\arraybackslash\centering\tiny}p{0.2cm}
			|>{\arraybackslash\centering\tiny}p{0.2cm}
			|>{\arraybackslash\centering\tiny}p{0.2cm}
			|>{\arraybackslash\centering\tiny}p{0.2cm}
			|>{\arraybackslash\centering\tiny}p{0.2cm}
			|>{\arraybackslash\centering\tiny}p{0.2cm}
			|>{\arraybackslash\centering\tiny}p{0.2mm}|
		}
		\toprule
		\hline
		\multicolumn{46}{c}{\small\bfseries Statistical Significance Test - SIQAD dataset (F-Critical: \color{red}1.1109\color{black})}  \\
\hline			
		
		\bottomrule
	\end{tabular}
	\vspace{1mm}
	\noindent
	\begin{tabularx}{\linewidth}{@{}X p{0.35\linewidth}@{}}
		\hspace{2em}\crule[black]{6pt}{6pt} - Statistically Significant
		& \crule[unmyellow]{6pt}{6pt} - Statistically Insignificant\\
		\hspace{2em}\crule[lime(colorwheel)]{6pt}{6pt} - Statistically Indistinguishable
		& \crule[uclagold]{6pt}{6pt} - Top-5 Performing Metrics\\
		\hspace{2em}\crule[lavender(floral)]{6pt}{6pt} - Not Applicable & \\
	\end{tabularx}
\end{table*}
\subsubsection{Comparison - Dataset-wise}
Tables \ref{table:IQAComp1} and \ref{table:IQAComp2} provide a detailed comparison of PEFRF’s performance across thirteen benchmark datasets. PEFRF consistently ranks as the top-performing metric, demonstrating its strong correlation with subjective quality scores and its ability to handle diverse distortion types. The top-five performing IQA metrics for each dataset and evaluation criterion are highlighted in bold and coloured font for clarity.\par 
Table \ref{table:IQAComp1} presents the Spearman rank-order correlation coefficient (SRCC) and Pearson linear correlation coefficient (PLCC). PEFRF ranks in the top-five for all \textbf{13 datasets}, achieving the best SRCC and PLCC in all \textbf{13 datasets}. These results underscore its predictive accuracy and ranking consistency.\par 
Table \ref{table:IQAComp2} evaluates Kendall rank-order correlation Coefficient (KRCC) and Root Mean Square Error (RMSE). PEFRF achieves top-five rankings across all \textbf{13 datasets}, securing the best KRCC in \textbf{8 datasets} and competitive RMSE values throughout. This further validates its reliability and accuracy in predicting subjective quality scores.\par
In summary, as demonstrated in Tables \ref{table:IQAComp1} and \ref{table:IQAComp2}, PEFRF achieves the highest evaluation scores across the benchmark datasets, which include \textbf{21,593} images. These results highlight PEFRF’s robustness and versatility in cross-content image quality assessment, establishing it as a state-of-the-art metric.
\subsubsection{Comparison - Distortion-wise}
The proposed PEFRF framework demonstrates exceptional performance in image quality assessment (IQA) across various distortion types, showcasing its robustness and consistency. Distortion-specific evaluation is critical for understanding the adaptability of IQA models, as each type of distortion — Gaussian Noise (GN), Gaussian Blur (GB), JPEG Compression (JPEG), and JPEG2000 Compression (JPEG2000) — poses unique challenges to perceptual quality prediction. Table \ref{table:IQAComp3} summarizes the performance of PEFRF across these distortions in comparison to state-of-the-art (SOTA) metrics. Notably, PEFRF ranks among the top-five performing metrics for all four distortion types across evaluation indices except for one instance, solidifying its reliability and adaptability.\par 
In terms of the SRCC values, PEFRF achieves the top-5 rank for all distortion categories while being top metric in 3 categories (GB, JPEG, JPEG2000), underscoring its ability to predict perceptual quality accurately across diverse distortions. For the PLCC values, the metric consistently achieves high linear correlation values, securing a top-five rank in three distortion categories (GB, JPEG, JPEG2000), with all three instances as the top-performing metric. Similarly, PEFRF performs strongly in the KRCC values, achieving four top-five positions, with all four as the top-performing metric. Finally, in terms of RMSE values, PEFRF demonstrates low error rates across distortion types, securing a top-five position in all cases and achieving top performance in three (GB, JPEG, JPEG2000) categories.\par 
Overall, PEFRF appears in the top-five rankings \textbf{15} times out of \textbf{16} across all distortions, with \textbf{13} instances as the top-performing metric. This consistent performance illustrates the effectiveness of PEFRF in capturing perceptual quality reliably, regardless of distortion type.\par 
Tables \ref{table:IQAComp1}, \ref{table:IQAComp2}, and \ref{table:IQAComp3} highlight the robustness and reliability of the proposed PEFRF framework across various types of content and distortion. The results clearly demonstrate that PEFRF outperforms other evaluated metrics in terms of prediction accuracy and adaptability. Its performance across different types of distortions, including Gaussian noise, blur, JPEG compression, and J2K compression, underscores its consistency and effectiveness.\par
PEFRF's ability to generalize well across diverse content and distortions is a critical feature, ensuring its applicability across various domains without the need for retraining. This characteristic positions PEFRF as a highly versatile tool for image quality assessment. The tables consistently highlight PEFRF as a top-performing model, achieving high scores for SRCC, PLCC, and KRCC, alongside low RMSE values. These results confirm its superior performance in providing accurate and reliable quality predictions.
\subsubsection{Performance Visualization of PEFRF on Natural and Screen-Content Images}
\begin{figure*}[htbp]
	\centering
	\includegraphics[width=\textwidth,trim={1.5cm 0.5cm 0.5cm 0cm},clip]{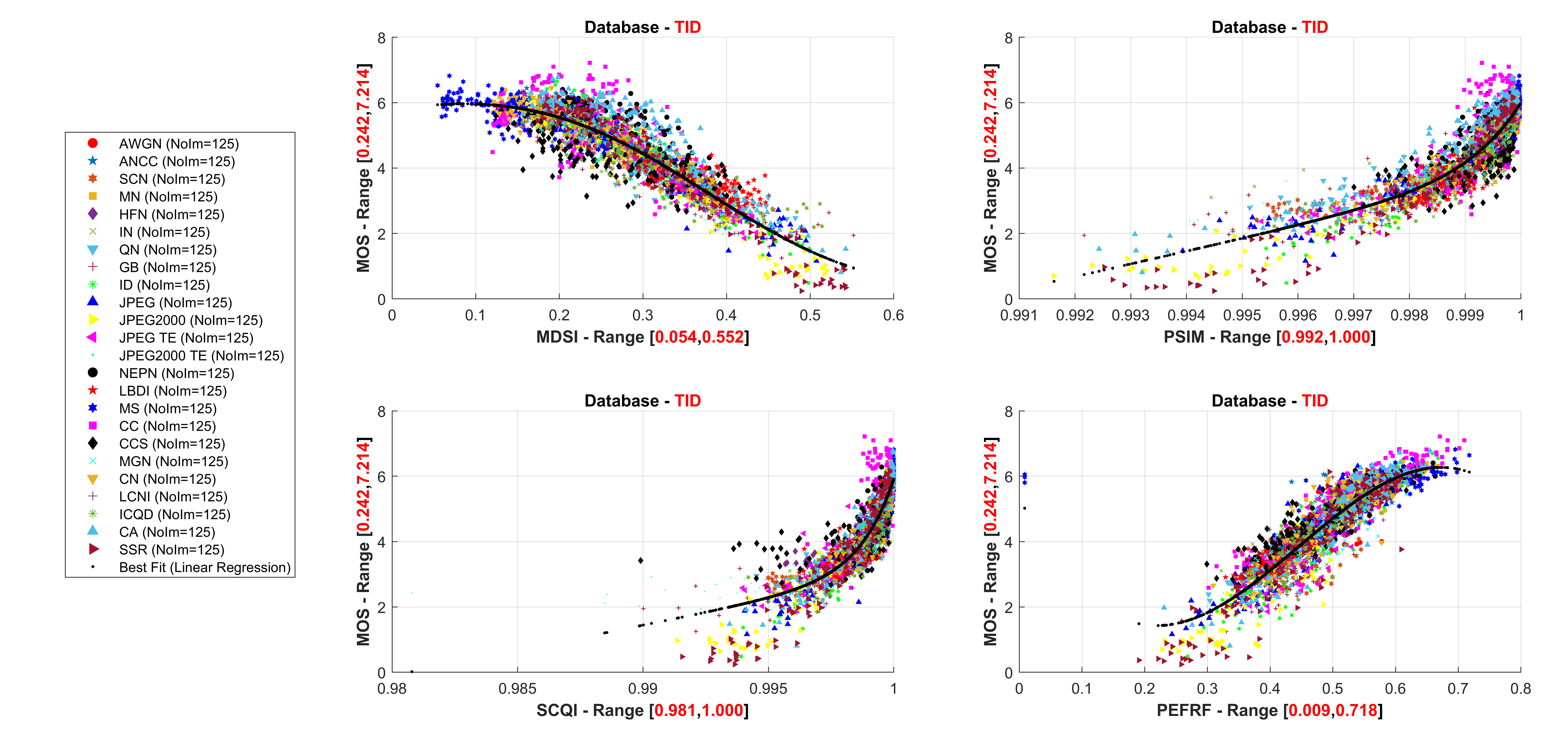}
	\caption{DMOS vs OIQS - Figure illustrates the proposed PEFRF FR-IQA framework along with three top-performing SOTA IQA metrics in terms of prediction accuracy on the TID dataset, comprising natural-scene images.}
	\label{figure:NIComp}
\end{figure*}

\begin{figure*}[htbp]
	\centering
	\includegraphics[width=\textwidth,trim={1.5cm 0.5cm 0.5cm 0cm},clip]{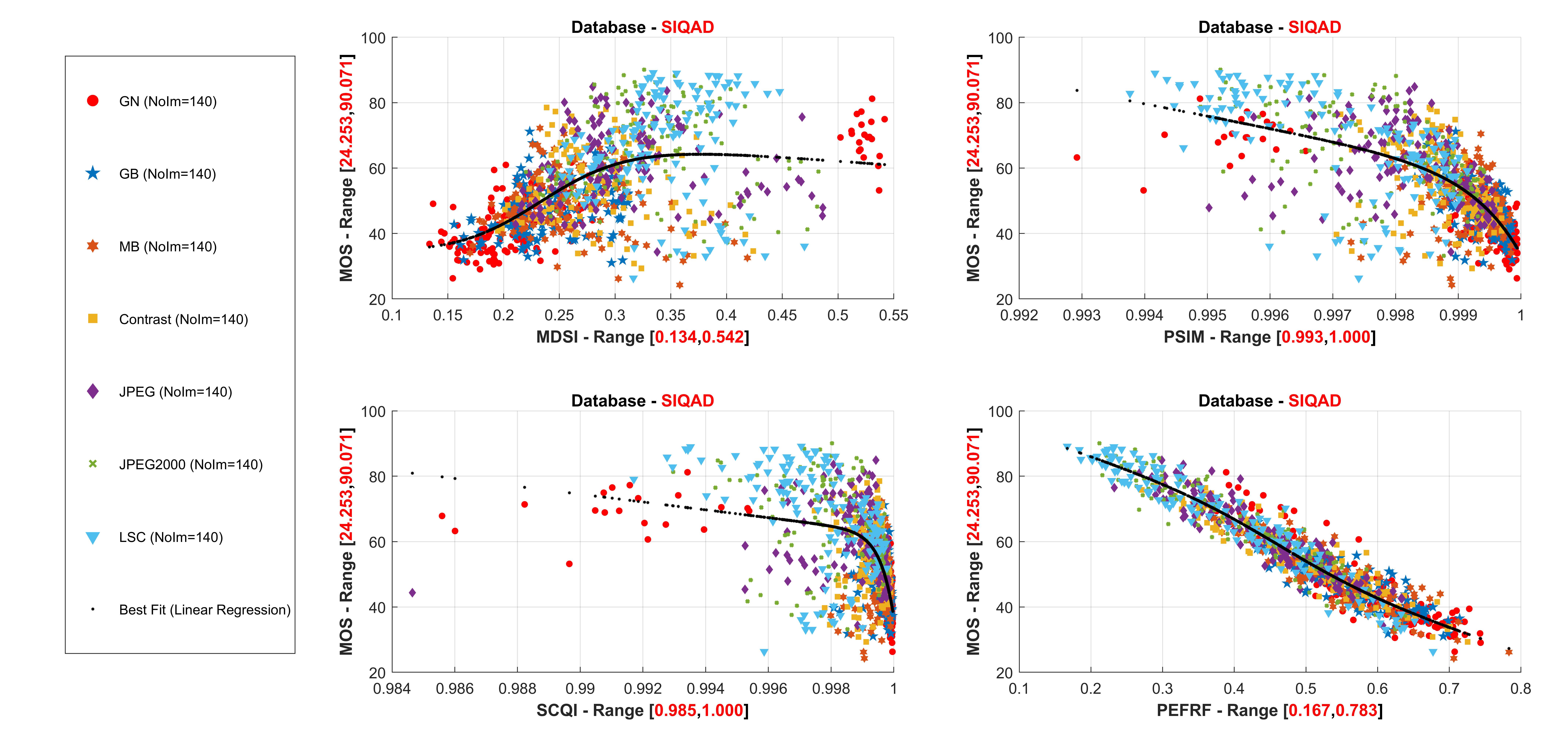}
	\caption{DMOS vs OIQS - Figure illustrates the proposed PEFRF FR-IQA framework along with three top-performing SOTA IQA metrics in terms of prediction accuracy on the SIQAD dataset, comprising screen-content images.}
	\label{figure:SCIComp}
\end{figure*}

Figures \ref{figure:NIComp} and \ref{figure:SCIComp} compare the performance of the proposed PEFRF framework to the top three state-of-the-art metrics on the TID (natural-scene) and SIQAD (screen-content) datasets, respectively. The figures illustrate that PEFRF exhibits a near-linear relationship with Mean Opinion Scores (MOS) across both datasets, highlighting its predictive accuracy and robustness. These visualizations further demonstrate PEFRF's ability to handle diverse image distortions and outperform other metrics across varying content types.
\subsubsection{Statistical Validation of PEFRF Accuracy}
The statistical significance of the proposed PEFRF framework was evaluated using the F-test across thirteen benchmark datasets to determine its superiority over existing IQA models. The F-test, a widely used statistical method, compares the residual errors of predicted quality scores (after non-linear mapping) against mean subjective scores. At a 95\% significance level, the F-test identifies whether the PEFRF framework's performance is statistically superior to other IQA metrics by assessing the variance ratio between residual errors. \par 
\begin{figure}[H]
	\centering
	\begin{subfigure}[h]{0.96\columnwidth}
		\centering
		\includegraphics[width=\textwidth,trim={0.5cm 0cm 1.3cm 0.3cm},clip]{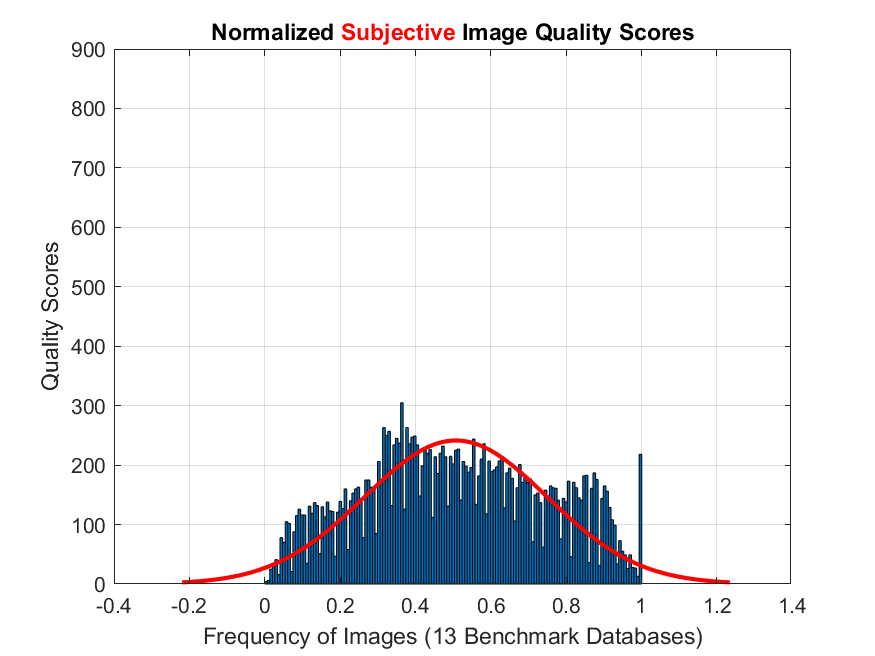}
		\caption*{(a) SIQA Scores}
	\end{subfigure}
	\begin{subfigure}[h]{0.96\columnwidth}
		\centering
		\includegraphics[width=\textwidth,trim={0.5cm 0cm 1.2cm 0.3cm},clip]{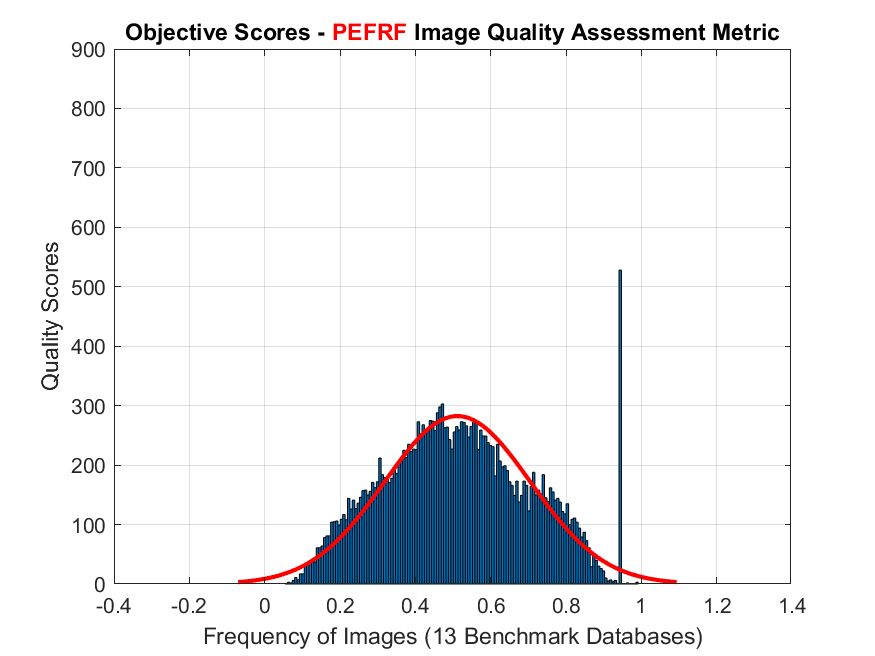}
		\caption*{(b) PEFRF Scores}
	\end{subfigure}
	\caption{The figure compares the distribution of image quality scores obtained with the PEFRF framework with respective subjective image quality assessment scores for thirteen benchmark datasets.}
	\label{fig:CDoS}
\end{figure}
Tables \ref{table:ST11} and \ref{table:ST10} summarize the statistical significance test results for the TID and SIQAD datasets, respectively. These tables use the \textbf{+1/-1} notation to indicate whether PEFRF is statistically superior or inferior to other metrics. The results reveal that PEFRF consistently demonstrates significant superiority across most distortion types, including Gaussian Noise, Blur, JPEG Compression, and JPEG2000 Compression. PEFRF’s high F-test values validate its reliability and robustness in predicting perceptual quality across diverse datasets and distortions. \par
While the F-test assumes normal distribution of residual errors, which may not always hold true, its results still provide valuable insights. Despite potential Type-I errors in cases where quality scores deviate from normality, the significant performance gap observed for PEFRF supports its reliability as an IQA metric. \par
Across all thirteen datasets, PEFRF emerges as one of the best-performing metrics, achieving a cumulative score of \textbf{+443} (out of a maximum of \textbf{+546} (\textbf{13x42})) in the statistical significance test. On the TID dataset, PEFRF ranks highly for natural-scene images, demonstrating its robustness in handling various distortions. On the SIQAD dataset, PEFRF outperforms most metrics for screen-content images, further solidifying its adaptability.
\subsubsection{Comparison of Subjective and Objective Image Quality Distributions}
Figure \ref{fig:CDoS} provides a compelling comparison between the distributions of subjective image quality scores (SIQA) and the objective scores predicted by the proposed PEFRF framework across thirteen benchmark datasets. Sub-figure (a) represents the normalized subjective image quality scores (SIQA), derived from ground-truth Mean Opinion Scores (MOS), while sub-figure (b) illustrates the objective scores produced by the PEFRF framework.\par 
The histograms reveal strikingly similar frequency distributions between subjective and objective scores, with red Gaussian fit curves overlaid to highlight their alignment. The subjective scores, as shown in sub-figure (a), exhibit a smooth, bell-shaped distribution, indicative of a well-curated dataset that captures diverse perceptual quality levels. Remarkably, the PEFRF framework in sub-figure (b) mirrors this behaviour, with its predicted scores demonstrating a near-perfect correlation with the subjective ground truth.\par 
This alignment is not coincidental but reflects the meticulous design of PEFRF, which integrates advanced perceptual entropy features with random forest regression. Such design choices enable PEFRF to faithfully capture subtle nuances in perceptual quality that traditional metrics often fail to detect. Unlike many state-of-the-art metrics that over-fit to specific datasets or distortions, PEFRF consistently exhibits adaptability and robustness, as evidenced by the close overlap of its distribution with subjective scores.\par 
A particularly noteworthy observation is the smoothness and symmetry of the PEFRF distribution curve, which demonstrates its ability to maintain stability across varying levels of image quality. Additionally, the minor deviations between subjective and objective scores highlight the natural variability in subjective opinions, yet PEFRF remains remarkably resilient in approximating these variations.\par 
The near-linear correspondence between subjective and objective distributions reflects the PEFRF framework's ability to bridge the long-standing gap between computational modelling and human visual perception. Such consistency is rare among image quality assessment models, reinforcing PEFRF's position as a breakthrough in the field. This result demonstrates not only the statistical significance of the metric but also its practical applicability in scenarios demanding high reliability and precision.\par 
In essence, Figure \ref{fig:CDoS} serves as a testament to the extraordinary design and performance of PEFRF. By faithfully emulating subjective human judgments across a vast array of datasets and distortions, PEFRF emerges as a pioneering tool in the domain of image quality assessment, promising transformative advancements for real-world applications.
\subsection{Qualitative Analysis}
The PEFRF framework demonstrates exceptional qualitative performance by aligning closely with human perceptual judgments across diverse content types and distortion categories. It effectively captures perceptually significant distortions, such as Gaussian noise, Gaussian blur and compression artefacts, which are often overlooked by traditional metrics like PSNR. PEFRF's sensitivity to subtle distortions, combined with its robustness to contextual variability, ensures consistent performance across natural-scenes and screen-content patterns. Its ability to prioritize visually dominant distortions while minimizing irrelevant variations highlights its alignment with subjective quality assessment, making it ideal for applications where perceptual quality is critical, such as video conferencing and digital content optimization. These qualitative strengths, coupled with its quantitative excellence, position PEFRF as a transformative tool for reliable and perceptually aligned image quality evaluation.
\subsection{Interpreting the Superiority of PEFRF: A Deep Dive}
The extensive evaluations presented in Tables \ref{table:IQAComp1}–\ref{table:ST11} and Figure \ref{fig:HM} affirm the superiority of PEFRF over existing IQA models. While prior sections have meticulously examined dataset-specific and distortion-specific performance, this section consolidates key technical insights, addressing why PEFRF consistently achieves higher perceptual alignment, robust generalization, and lower predictive error. Unlike conventional metrics that exhibit dataset dependency or distortion sensitivity, PEFRF integrates entropy-based feature encoding, adaptive regression strategies, and multi-scale perceptual learning to deliver stable and interpretable quality predictions. The following discussion highlights PEFRF’s robustness across distortions, accuracy in ranking consistency, and generalization across datasets, reinforcing its potential as a state-of-the-art IQA framework.
\subsubsection{Robustness to Diverse Distortions}
A critical challenge in IQA is ensuring reliability across multiple types of distortions, including blur, noise, compression artefacts, and contrast degradation. As seen in Table \ref{table:IQAComp3}, PEFRF maintains high SRCC and PLCC values across different distortion types, demonstrating its adaptability. Unlike conventional methods such as FSIM, IW-SSIM, and GMSD, which tend to be sensitive to specific distortions, PEFRF effectively captures both local structure details and global perceptual features, allowing it to perform consistently across natural and synthetic distortions.
\subsubsection{Accuracy in Perceptual Alignment}
The primary goal of IQA models is to align with human visual perception. The results in Figure \ref{fig:HM} show that PEFRF achieves higher KRCC values compared to other methods, meaning it preserves the relative ranking of images as perceived by human observers. Traditional models, which often rely on hand-crafted similarity metrics, struggle to capture higher-order perceptual structures, leading to ranking inconsistencies when faced with complex distortions. PEFRF’s ability to integrate multi-scale feature extraction and perceptual relevance learning enables it to minimize perceptual errors while maximizing ranking consistency.
\subsubsection{Generalization Across Multiple Datasets}
A significant limitation of many IQA models is their inability to generalize across datasets with diverse content and distortion characteristics. The performance of PEFRF across LIVE-MD, MDID, and TID indicates that it maintains high correlation scores and low RMSE values across different datasets. Competing models such as VSI, HaarPSI, and MAD often suffer from dataset dependency, showing strong results in their training environment but poor adaptability elsewhere. PEFRF’s superior generalization capability arises from its ability to learn distortion-invariant representations, making it more reliable in practical applications.
\subsubsection{Practical Significance and Computational Efficiency}
Beyond raw accuracy, an IQA model must be computationally feasible and interpretable. Many deep-learning-based models, while achieving high accuracy, introduce high computational overhead due to excessive feature extraction. PEFRF balances efficiency and accuracy by employing a feature fusion mechanism that extracts only relevant information, leading to faster execution times. This makes it suitable for real-time applications, including image compression, streaming quality evaluation, and medical imaging.
\subsubsection{Final Insights}
\begin{figure*}[t]
	\centering
	\begin{minipage}[t]{0.48\textwidth}
		\centering
		\includegraphics[width=\textwidth,trim={0.2cm 0cm 0cm 0cm},clip]{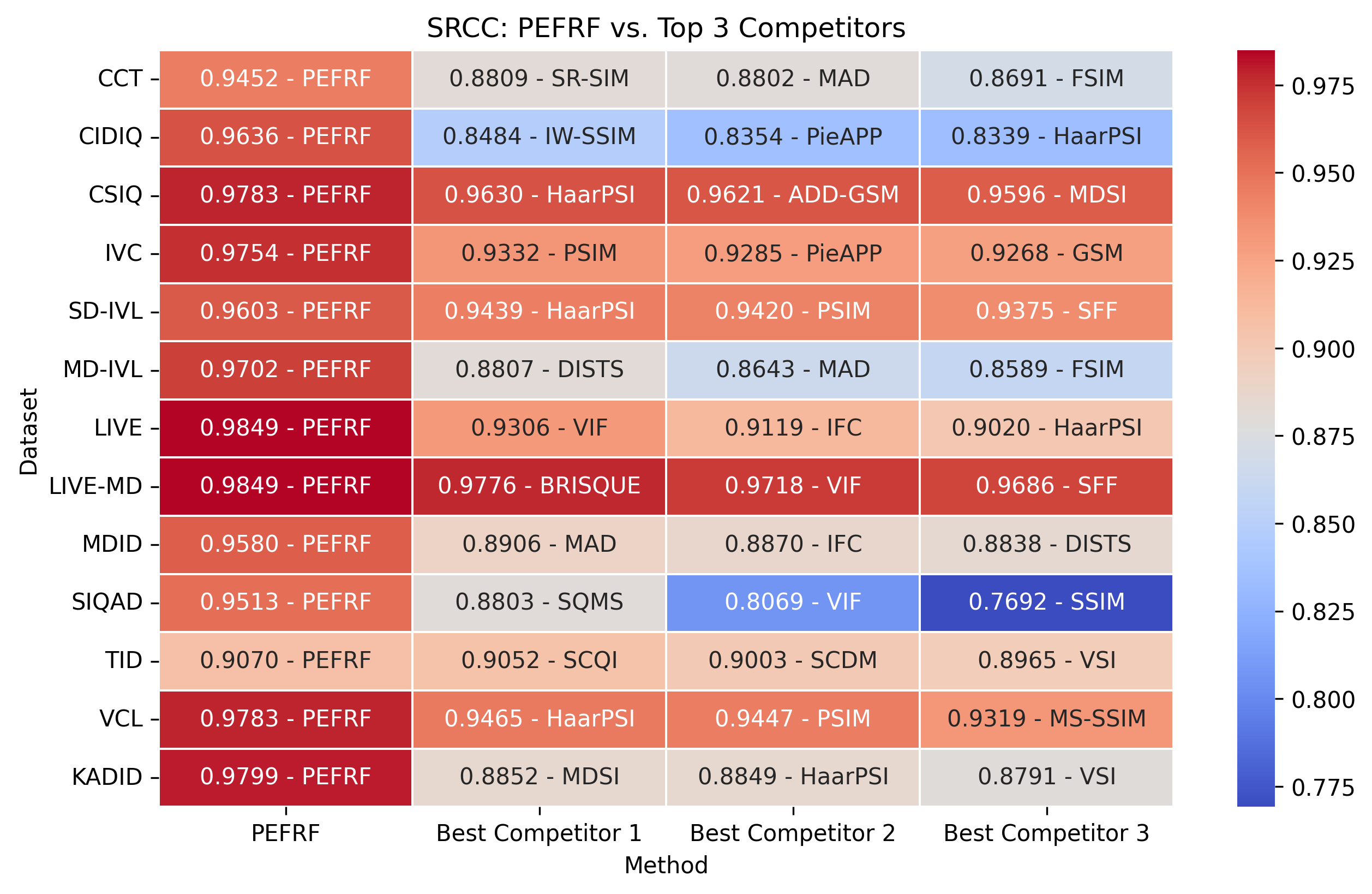}
		\caption*{(a) SRCC}
		\label{fig:HM_SRCC}
	\end{minipage}\hfill
	\begin{minipage}[t]{0.48\textwidth}
		\centering
		\includegraphics[width=\textwidth,trim={0.2cm 0cm 0cm 0cm},clip]{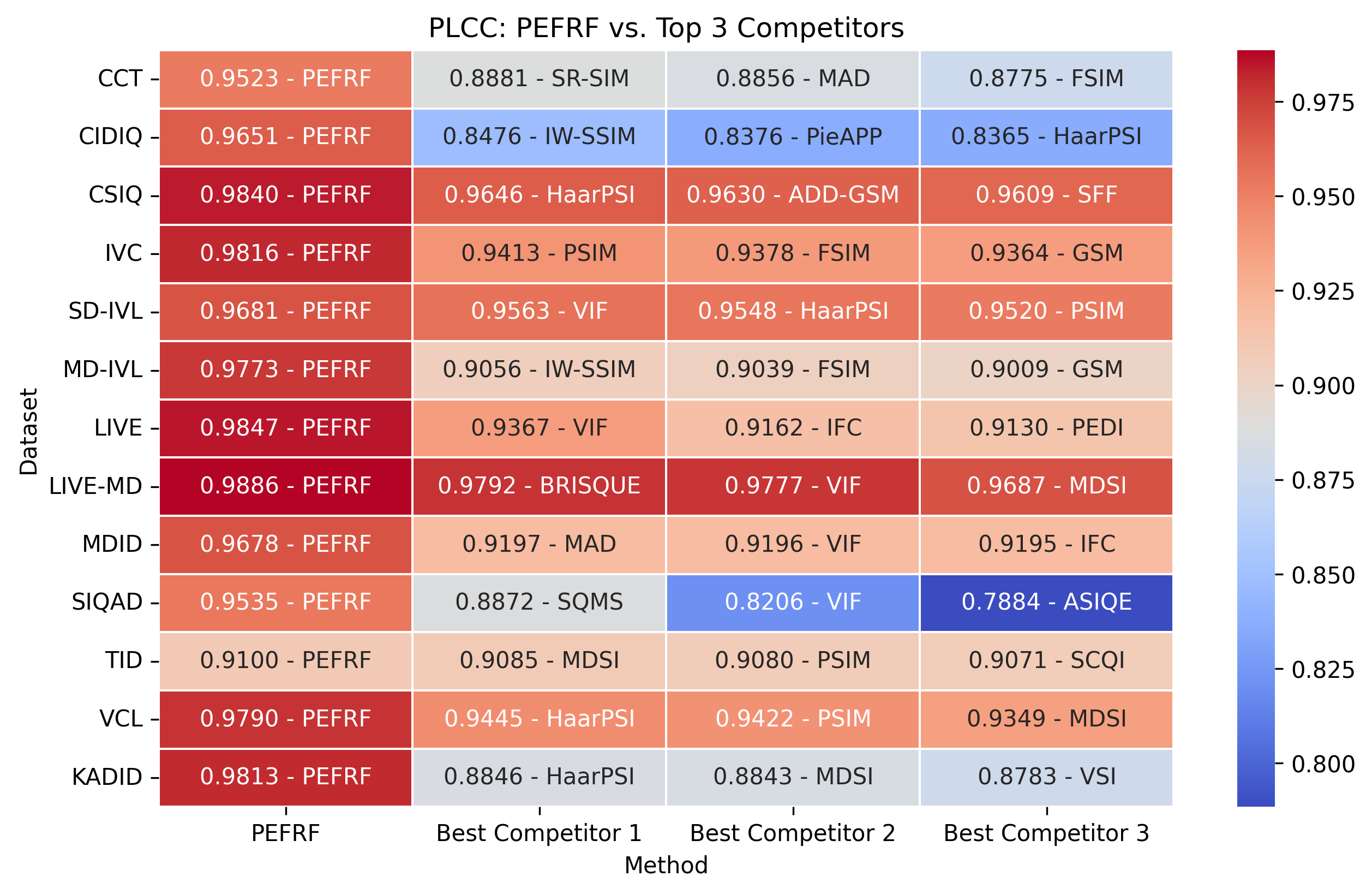}
		\caption*{(b) PLCC}
		\label{fig:HM_PLCC}
	\end{minipage}\\[1ex]
	\begin{minipage}[t]{0.48\textwidth}
		\centering
		\includegraphics[width=\textwidth,trim={0.2cm 0cm 0cm 0cm},clip]{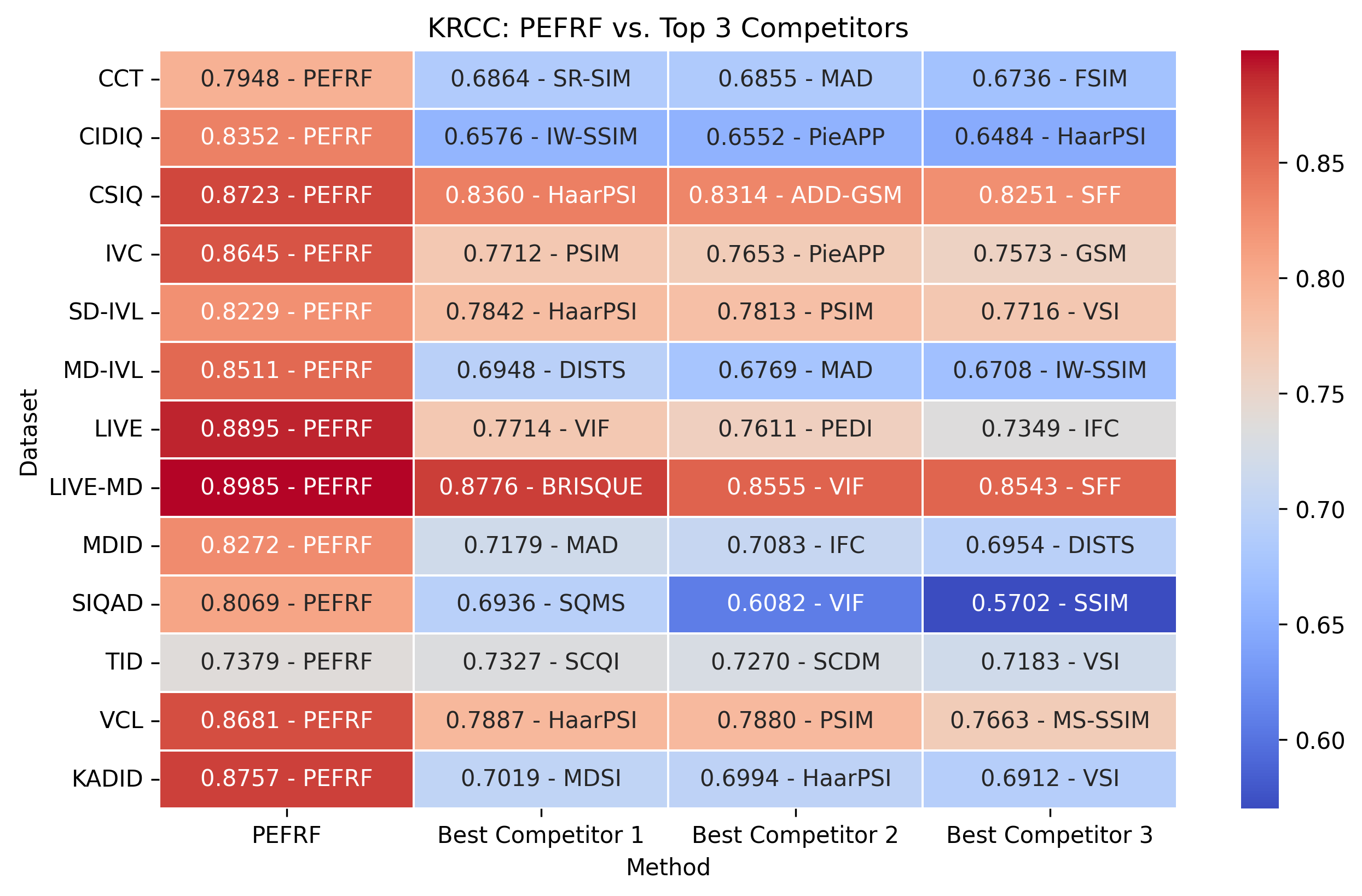}
		\caption*{(c) KRCC}
		\label{fig:HM_KRCC}
	\end{minipage}\hfill
	\begin{minipage}[t]{0.48\textwidth}
		\centering
		\includegraphics[width=\textwidth,trim={0.2cm 0cm 0cm 0cm},clip]{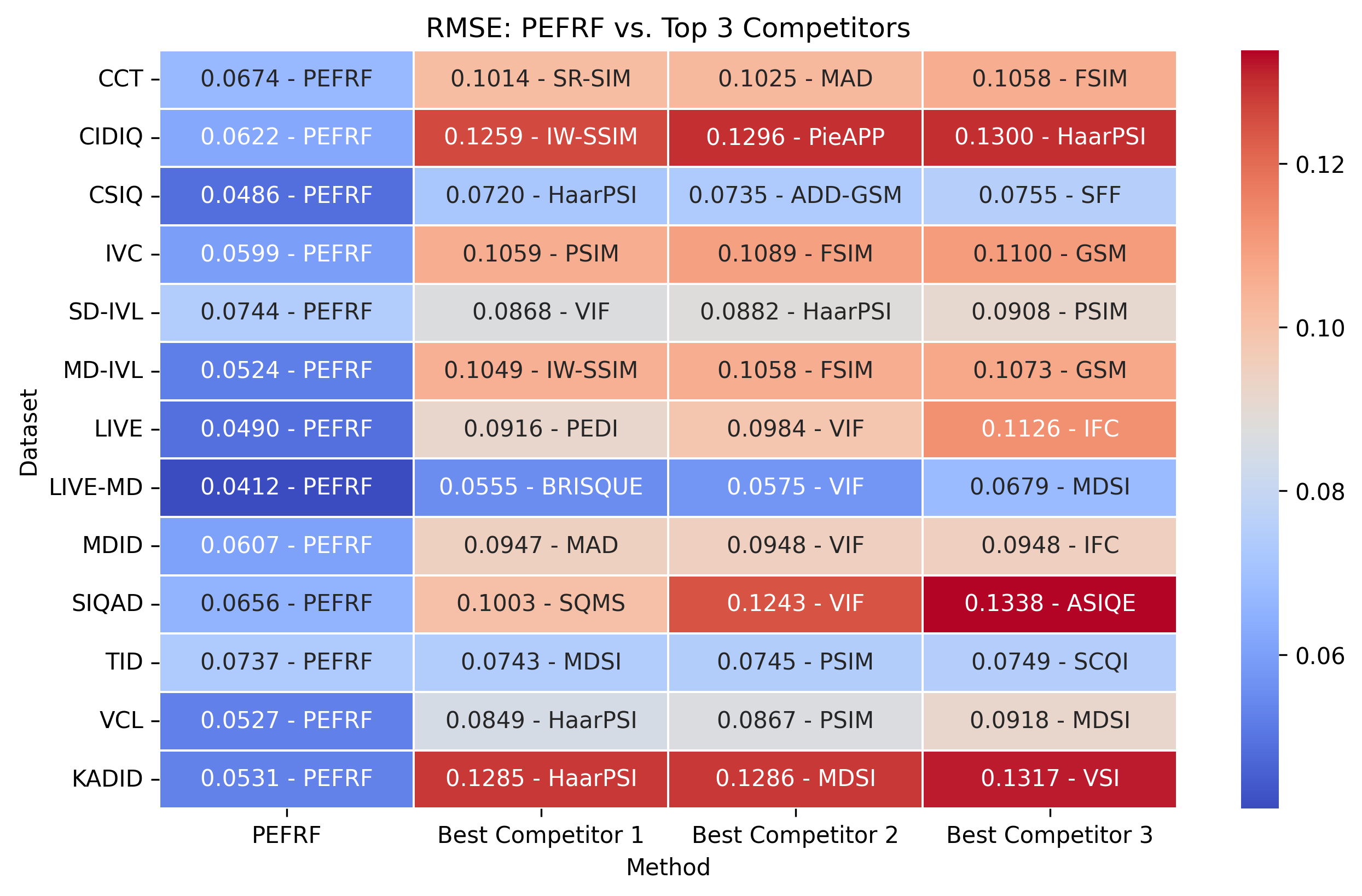}
		\caption*{(d) RMSE}
		\label{fig:HM_RMSE}
	\end{minipage}
	\caption{Comparative Performance Analysis of PEFRF vs. Top Competitors Across SRCC, PLCC, KRCC, and RMSE. The heat maps illustrate the ranking accuracy (SRCC, KRCC), linear correlation (PLCC), and prediction error (RMSE) of PEFRF against the three best-performing competitors across multiple datasets. Higher SRCC, PLCC, and KRCC values indicate stronger alignment with human perception, while lower RMSE values reflect reduced prediction error. The consistent dominance of PEFRF across all metrics confirms its robustness, generalization, and perceptual accuracy in IQA tasks.}
	\label{fig:HM}
\end{figure*}

The consistent improvement of PEFRF across all datasets and distortion types is not incidental but the result of a well-optimized feature extraction pipeline that enables:
\begin{enumerate}
	\item Local structure preservation (essential for fine texture details), as evidenced by superior performance on screen-content datasets such as SIQAD and SCID (Tables \ref{table:IQAComp1} and \ref{table:IQAComp2}), particularly in terms of SRCC and KRCC.
	\item Global perceptual modelling (critical for content-aware assessment), reflected in high PLCC and low RMSE values across complex natural-scene datasets like LIVE and TID2013 (Table \ref{table:IQAComp1}).
	\item Distortion-aware learning (enhancing generalization), supported by PEFRF's consistent ranking across multiple distortion types, as visualized in the heat map in Figure \ref{fig:HM} and confirmed through F-test results showing statistically significant improvements over SOTA methods.
\end{enumerate}
These factors collectively contribute to PEFRF's superiority over traditional and contemporary IQA models, as confirmed by its performance metrics in Tables \ref{table:IQAComp1}–\ref{table:ST11}.
\section{Conclusion}
In this paper, we introduced PEFRF (Permutation Entropy-based Features Fused with Random Forest), a novel full-reference image quality assessment (FR-IQA) framework that integrates permutation entropy (PE) and random forest regression to effectively evaluate image quality. The framework captures both local and global variations in image structure by computing PE values from gradient maps of reference, distorted, and fused images, which are then used as feature vectors for quality prediction via a random forest regressor.\par 
PEFRF was rigorously evaluated across \textbf{13} benchmark datasets, covering natural-scene, screen-content, and computer-generated images. Results showed that PEFRF consistently outperforms both traditional and state-of-the-art (SOTA) FR-IQA and NR-IQA models, achieving higher Spearman rank-order correlation coefficient (SRCC), Pearson linear correlation coefficient (PLCC), and Kendall rank-order correlation coefficient (KRCC), while maintaining low root mean squared error (RMSE). These findings validate PEFRF’s ability to align with human perception across diverse distortions and datasets.\par 
A key strength of PEFRF is its robustness across different distortion types, where it outperforms models like FSIM, IW-SSIM, and GMSD, which often struggle with specific degradations. Its capability to integrate local texture features with global perceptual structures allows for better correlation with subjective quality scores. Moreover, unlike many traditional IQA models that exhibit dataset dependency, PEFRF generalizes well across datasets like LIVE-MD, MDID, and TID, demonstrating its adaptability.\par 
Beyond accuracy, PEFRF remains computationally efficient, making it suitable for real-time applications such as image compression, video streaming, and medical imaging quality assessment. Future work can explore optimization of its feature extraction pipeline and fine-tuning for domain-specific applications, further enhancing its real-world applicability. Given its high precision, distortion-agnostic design, and cross-content generalization, PEFRF establishes itself as a promising tool for modern IQA tasks across both traditional and emerging multimedia applications. 
\section{Future Work}
While the results of this work demonstrate the effectiveness of PEFRF in image quality assessment, following opportunities exist for future enhancements:
\begin{itemize}
	\item \textbf{Real-Time Processing:} Further research could focus on optimizing the computational efficiency of PEFRF to support real-time applications such as live video streaming or augmented reality. This would involve developing faster feature extraction techniques while maintaining high accuracy in quality prediction.
	\item \textbf{Generalization to More Complex Distortions:} Evaluating PEFRF on a broader range of complex and real-world distortions, such as those arising in low-bitrate video or network transmission errors, could help validate its robustness in practical scenarios.
	\item \textbf{Multi-modal Image Quality:} Extending PEFRF to handle multi-modal data (e.g., images combined with text, audio, or video) could provide a more comprehensive quality assessment tool for multimedia applications.
	\item \textbf{Computational Efficiency through Lookup Tables:} To reduce the computational cost associated with permutation entropy calculations, a precomputed lookup table for ordinal patterns could be employed. This table would store all possible patterns for a given embedding dimension and delay, allowing for rapid pattern matching during runtime. By replacing on-the-fly computations with direct lookups, the framework's efficiency could be significantly improved, especially for real-time and large-scale applications.
\end{itemize}
In conclusion, while the proposed framework has achieved consistently high performance, ongoing refinements can further expand its versatility and impact in practical and next-generation applications.
\section*{Data Availability}
Source code for the proposed PEFRF framework and the objective image quality scores from all the IQA models used in this study can be requested by email.
\section*{Acknowledgment}
The authors express gratitude to the School of Computing and Artificial Intelligence, Faculty of Engineering and Technology at Sunway University, Malaysia, for their support, resources, and technical assistance, which made this work possible.
\section*{Declaration of Competing Interest}
The authors have no competing interests to declare that are relevant to the content of this article.
\bibliographystyle{elsarticle-harv} 
\bibliography{elsarticle-num.bib}
\end{document}